# Epidemiology, Trajectories and Outcomes of Acute Kidney Injury Among Hospitalized Patients: A Retrospective Multicenter Large Cohort Study

Esra Adiyeke, PhD[a,b], Yuanfang Ren, PhD[a,b], Shmuel Fogel, BS[a,b], Parisa Rashidi, PhD[a,c], Mark Segal, MD PhD [b], Elizabeth A. Shenkman, PhD[d], Azra Bihorac, MD, MS[a,b], Tezcan Ozrazgat-Baslanti, PhD[a,b]

[a] Intelligent Clinical Care Center, University of Florida, Gainesville, FL

[b] Department of Medicine, Division of Nephrology, Hypertension, and Renal Transplantation, University of Florida, Gainesville, FL.

[c] Department of Biomedical Engineering, University of Florida, Gainesville, FL.

[d] Department of Health Outcomes and Biomedical Informatics, University of Florida, Gainesville, FL.

**Corresponding author:** Tezcan Ozrazgat-Baslanti PhD, Department of Medicine, Division of Nephrology, Hypertension, and Renal Transplantation, PO Box 100224, Gainesville, FL 32610-0224. Telephone: (352) 294-8580; Fax: (352) 392-5465; Email: tezcan@ufl.edu

**ABSTRACT**

**Rationale & Objective:** Acute kidney injury (AKI) is a clinical syndrome affecting almost one fifth of hospitalized patients, as well as more than half of the patients who are admitted to the intensive care unit (ICU). Stratifying AKI patients into groups based on severity and duration would facilitate more targeted efforts for treating AKI. We analyzed the impact of AKI trajectories which are rapidly reversed AKI, persistent AKI with renal recovery, and persistent AKI without renal recovery on patients' clinical outcomes.

**Study Design:** Retrospective, multicenter and longitudinal cohort study

**Setting & Participants:** 935,679 patients who were admitted between 2012 and 2020 to health centers included in OneFlorida+ Network

**Exposure(s):** AKI trajectories

**Outcomes:** Hospital, 30-day, 1-year, and 3-year mortality, kidney replacement therapy, new chronic kidney disease (CKD) within 90 days or 1-year of discharge, CKD progression within 1-year of discharge, resource utilization, hospital disposition, and major complications during hospitalization.

**Analytical Approach:** Kaplan-Meier estimators and survival curves, Cox proportional-hazards regression model, logistic regression model, Kruskal-Wallis test, analysis of variance, chi-square, Fisher's exact test.

**Results:** Among 2,187,254 encounters, 14% had AKI, of which 63%, 21%, and 16% had Stage 1, 2, and 3, respectively, as the worst AKI stage. Fraction of patients with persistent AKI was 31%. Patients with AKI had worse clinical outcomes and increased resource utilization compared to patients without the condition. One-year mortality was 5 times greater for patients with persistent AKI compared to those without AKI.

**Limitations:** Serum creatinine criteria used, ICU in and out times were not available, only Florida based patients

**Conclusions:** Persistent AKI was associated with prolonged hospitalization, increased ICU admission and mortality compared to the other groups. This may emphasize the critical need for devising strategies targeting effective management of AKI and prevention of persisting AKI.

## Introduction

Acute kidney injury (AKI) is a prevalent clinical syndrome affecting over 20% of hospitalized patients and up to 60% of critical care patients.[1-5] Prior research on AKI has demonstrated that it is associated with a significantly increased risk of developing chronic kidney disease (CKD) and cardiovascular disease, as well as increased mortality.[6-11] Currently, AKI is treated as a syndrome with supportive care as opposed to targeted therapy aimed at the underlying cause, and new research seeks to stratify patients with AKI into phenotypes that can receive more specific treatments.[12, 13] Additionally, given the long-term consequences of AKI, an emphasis has been placed on encouraging patients to follow up with a nephrologist following hospital discharge.[11]

Stratifying AKI patients into groups with more nuanced characterizations such as severity and duration, would facilitate more targeted efforts for treating AKI. While the previous lack of uniform criteria for diagnosing and staging AKI hindered this, it is now feasible thanks to an accumulation of electronic health records (EHR) and recently developed, clear staging and persistence criteria standardized by the Kidney Disease Improving Global Guidelines (KDIGO) and Acute Disease Quality Initiative (ADQI) guidelines.[14, 15] Owing to these advancements, several studies have described AKI prevalence, trajectory, mortality, and biomarker profiles among patients receiving different level of care, such as for hospital admissions[8, 16] or admissions exclusive to critical care[6, 17], with various etiologies including sepsis[18] and surgery[19] and subgroups like coronavirus disease 2019 (COVID-19)[20]. Despite their important merits, these studies have not investigated impact of AKI on long-term clinical outcomes, have not considered AKI's duration aspect, or were based on a single-center design. In this study, we retrospectively analyzed the impact of AKI trajectories on patients' clinical outcomes, including resource use, and assessed the relative importance of AKI severity, persistence and renal recovery on patients' long-term survival in a large, multicenter cohort.

We performed our analyses on a cohort derived from the OneFlorida+ Clinical Research Network, which is a clinical consortium of 14 health systems that serve almost 20 million patients [21, 22]. This large and multicenter cohort setting strengthens our study with its more diverse patient sampling from both urban and rural areas, aiding generalizable inferences compared to studies based on a single center with similar size. Another contribution  of our study is that is considers persistent AKI characterization, which is a relatively novel AKI taxonomy that requires more in-depth investigations compared to existing research on AKI.[23] Considering the aforementioned limitations in previous studies, we aim to fill this gap by investigating the impact of AKI persistence and renal recovery on both short- and long-term outcomes in a diverse set of hospitalized patients.

## Methods

### *Study design, cohort and clinical assessments*

In this retrospective, observational study, we used EHR data from the OneFlorida+ network, which is a partner of National Patient Centered Clinical Research Network (PCORnet).[21] The OneFlorida Data Trust is a research data repository initiated and maintained by OneFlorida+; it assembles patients' EHR along with claims data, death data, tumor data and external datasets that are publicly available. The OneFlorida+ dataset contains patient information, including demographics, enrollment status, vital signs, conditions, encounters, diagnoses, procedures, medications, and lab results.

We received longitudinal EHR data for 4,140,280 encounters from 1,260,000 patients admitted to healthcare institutions that are part of OneFlorida+ network, between January 1, 2012 and July 21, 2020. We excluded patients who were less than 18 years old as of their admission date, as well as patients with no serum creatinine record during their hospital stay and within first 48 hours of hospital admission. After excluding these subjects, we obtained our study cohort of 2,187,254 encounters from 935,679 patients (Supplemental Figure 1). This study was approved by the Institutional Review Board of the University of Florida and the University of Florida Privacy Office (IRB201902576).

To determine patients' comorbidities, we utilized International Classification of Diseases, Ninth Revision (ICD-9) and International Classification of Diseases Tenth Revision (ICD-10) codes capped with 50 elements at most. We quantified patients' chronic disease burden by computing the Charlson Comorbidity Index (CCI) score with patients' diagnosis records spanning past 365 days from their admission.[24] We identified mechanical ventilation therapy by considering ICD 9 and ICD 10 codes (96.7, 96.70, 96.71, 96.72, 5A1935Z, 5A1945Z, 5A1955Z, 94657, 94656, 1015098, 94003, 94002, 94004, 1014859) in addition to Current Procedural Terminology (CPT) codes (V46.1, V46.14, Z99.11, Z99.12, J95.85, 95.859). Similarly, to identify intensive care unit (ICU) patients, we utilized CPT codes listed as 99291, 99292, 0188T and 0189T. The reported number of nephrotoxic drug groups two days following AKI onset were based on aminoglycosides, diuretics, vancomycin, angiotensin converting enzyme (ACE) inhibitor or angiotensin receptor blockers (ARB), non-steroidal anti-inflammatory drugs (NSAIDs), and vasopressors and/or inotropes.

***Diagnostic criteria***

We utilized computable phenotyping algorithms that we previously developed and validated to identify the presence and staging of CKD and AKI, as well as the trajectory of AKI based on KDIGO standardized serum creatinine criteria.[25] As per KDIGO serum creatinine criteria, presence of AKI was defined as an increase from the reference creatinine of ≥ 0.3 within 48 hours or ≥ 1.5-fold increase within seven days. Stage 1 AKI was defined by elevation in serum creatinine ≥ 0.3 or a 1.5–1.9-fold increase above the patient’s baseline serum creatinine levels. Stage 2 AKI was defined by a 2–3-fold increase from baseline creatinine levels, and Stage 3 AKI was defined by a greater than 3-fold increase. Reference serum creatinine values were determined by either considering the preadmission records or estimating with the Chronic Kidney Disease Epidemiology Collaboration (CKD-EPI) study refit without the race multiplier formula, with an assumed baseline glomerular filtration rate value of 75 mL/min/per 1.73 $m^2$.

We classified a patient’s AKI stage based on their most severe stage during hospitalization. AKI trajectory was defined as “rapidly reversed” if it resolved within 48 hours and “persistent” otherwise. Stage 1 AKI was considered mild AKI, whereas Stage 2 and Stage 3 AKI were considered to be severe. For time to death analysis, we created subgroups obtained by AKI trajectory and severity group combinations. We presented G-stage CKD by using the estimated glomerular filtration rate (eGFR), which was calculated with a CKD-EPI formula. We refer to Ozrazgat-Baslanti et al. (2019) for details on estimating reference creatinine and descriptions of AKI algorithm pipeline along with relevant entities including Logical Observation Identifiers Names and Codes (LOINC), ICD-9, ICD-10 and Current Procedural Terminology (CPT) codes utilized for end-stage kidney disease (ESKD), kidney replacement therapy (KRT), CKD, CKD progress and AKI identification.[25]

***Outcomes and Analyses***

Primary clinical outcomes that we assessed were hospital, 30-day, 1-year, and 3-year mortality while renal outcomes of interest were KRT, new CKD within 90 days or 1-year of discharge, and CKD progression within 1-year of discharge. Additional outcomes measured include resource utilization, hospital disposition, and several major complications during hospitalization.

We evaluated the survival outcomes of each trajectory class in 935,679 patients with Kaplan-Meier estimators and compared survival curves via a log-rank test. To obtain adjusted Kaplan-Meier curves, we calculated propensity score-based inverse weights, where the probability of developing a trajectory class was computed with a multinomial logistic regression model in which patient age (which was dichotomized by being older than 65 years or not), sex, race (which was dichotomized by being African-American or not) and chronic disease burden (indicated with CCI score). In evaluating associations among AKI entities (AKI vs no AKI, AKI severity, AKI trajectories and AKI trajectories coupled with severity) and time to death, we generated Cox proportional-hazards regression models while adjusting for age, sex, race, CCI score, need for mechanical ventilation, and need for ICU admission, while analyzing all cohorts and ICU subgroups. For sub-group analyses of time to death for patients who had not been admitted to ICU, we adjusted for age, sex, race, and CCI score.

To model hospital mortality, we used multivariate logistic regression and included patient demographics and CCI. We analyzed these derived models to investigate the difference in association by extending variable set with variables indicating being severe AKI or Stage 3 AKI group. We modeled patient survivals starting from their hospital discharge date and up to three years. Harrell's concordance index was calculated in reporting model discrimination. For continuous variables Kruskal-Wallis test or analysis of variance, and for categorical variables, chi-square or Fisher's exact test was used to test the differences between groups, as appropriate. We reported p-values adjusted with Bonferroni methods for multiple comparisons. All statistical analyses were performed using R V.4.1.2 and Python V.3.8 software.

## Results

### *Cohort Characteristics*

Our study cohort had an average age of 59 (standard deviation [SD], 22) with 53% (n = 1,167,105) of cohort involving females (Table 1). Chronic pulmonary disease was the comorbidity with highest prevalence (28%) while diseases related to the circulatory system (43%) were found to be the most frequent admission diagnosis (Supplemental Table 1). Six percent of all patient encounters had surgery during their hospitalization, while 8% were transferred from another institution. Among overall patient encounters, 4% (n = 83,478) were admitted to ICU. Average age for ICU (mean 59, SD 21) and non-ICU cohorts were similar (mean 59, SD 22) while the percentage of

**Table 1.** Patient characteristics stratified by trajectories of AKI in all cohort.

| **Variables** | **All Cohort (N= 2,187,254)** | **AKI (N= 301,464, %14)** | **Persistent AKI without renal recovery (N= 92,607, %4)** | **Persistent AKI with renal recovery (N=45,749, %2)** | **Rapidly reversed AKI (N=163,108, %8)** | **No AKI (N= 1,885,790, %86)** |
|---|---|---|---|---|---|---|
| **Preadmission Clinical Characteristics** | | | | | | |
| Age, years, mean (SD) | 59 (22) | 69 (24)[a] | 74 (27)[abc] | 68 (22)[ab] | 67 (23)[a] | 58 (22) |
| Female sex, n (%) | 1,167,105 (53) | 157,121 (52)[a] | 51,895 (56)[abc] | 23,035 (50)[a] | 82,191 (50)[a] | 1,009,984 (54) |
| African American ethnicity, n (%) | 432,591 (20) | 65,225 (22)[a] | 19,048 (21)[abc] | 10,133 (22)[a] | 36,044 (22)[a] | 367,366 (19) |
| Transfer from another hospital | 172,607 (8) | 34,705 (12)[a] | 12,792 (14)[abc] | 6,618 (14)[ab] | 15,295 (9)[a] | 137,902 (7) |
| **Comorbidities, n (%)** | | | | | | |
| Hypertension | 362,545 (17) | 89,852 (30)[a] | 24,840 (27)[abc] | 16,053 (35)[ab] | 48,959 (30)[a] | 272,693 (14) |
| Chronic pulmonary disease | 604,472 (28) | 96,967 (32)[a] | 26,740 (29)[abc] | 15,760 (34)[ab] | 54,467 (33)[a] | 507,505 (27) |
| Cardiovascular disease [d] | 562,903 (26) | 115,883 (38)[a] | 32,766 (35)[abc] | 19,990 (44)[ab] | 63,127 (39)[a] | 447,020 (24) |
| Diabetes mellitus | 599,788 (27) | 113,184 (38)[a] | 30,730 (33)[abc] | 18,134 (40)[a] | 64,320 (39)[a] | 486,604 (26) |
| Chronic kidney disease | 475,640 (22) | 111,501 (37)[a] | 27,982 (30)[abc] | 21,240 (46)[ab] | 62,279 (38)[a] | 364,139 (19) |
| Moderate/Severe (≥ G-Stage 3) | 263,644 (55) | 64,829 (58) [a] | 15,022 (54)[abc] | 12,716 (60)[a] | 37,091 (60)[a] | 198,815 (55) |
| Preadmission estimated glomerular filtration rate (mL/min per 1.73 $m^2$), median (IQR) | 56.59 (40.87, 74.76) | 54.24 (37.75, 74.35)[a] | 57.03 (38.45, 81.69)[bc] | 53.04 (36.75, 72.74)[ab] | 53.71 (37.88, 72.30)[a] | 57.24 (41.76, 74.87) |
| **Kidney function within 48 hours of the admission** | | | | | | |
| AKI, n (%) | | | | | | |
| Stage 1 | 138,831 (6) | 138,831 (46)[a] | 38,182 (41)[abc] | 20,979 (46)[ab] | 131,083 (80)[a] | 0 (0) |
| Stage 2 | 45,382 (2) | 45,382 (15)[a] | 27,944 (30)[abc] | 13,226 (29)[ab] | 22,349 (14)[a] | 0 (0) |
| Stage 3 | | | | | | |
| Stage 3 without RRT | 29,674 (1) | 29,674 (10)[a] | 24,935 (27)[ab] | 11,124 (24)[ab] | 9,648 (6)[a] | 0 (0) |

| Variables | All Cohort (N= 2,187,254) | AKI (N= 301,464, %14) | Persistent AKI without renal recovery (N= 92,607, %4) | Persistent AKI with renal recovery (N=45,749, %2) | Rapidly reversed AKI (N=163,108, %8) | No AKI (N= 1,885,790, %86) |
|---|---|---|---|---|---|---|
| Stage 3 with RRT | 529 (0) | 529 (0)[a] | 1546 (2)[abc] | 420 (1)[ab] | 28 (0)[a] | 0 (0) |
| Highest blood urea nitrogen (mg/dl), mean (SD) | 20 (14) | 33 (23)[a] | 35 (26)[abc] | 40 (27)[ab] | 30 (19)[a] | 17 (11) |
| Highest serum creatinine (mg/dl), median (IQR) | 0.92 (0.74, 1.20) | 1.50 (1.09, 2.17)[a] | 1.51 (1.06, 2.40)[abc] | 1.80 (1.21, 2.75)[ab] | 1.44 (1.08, 1.96)[a] | 0.89 (0.71, 1.10) |
| Reference creatinine (mg/dl), median (IQR) | 0.82 (0.67, 1.01) | 0.88 (0.68, 1.15)[a] | 0.80 (0.58, 1.06)[abc] | 0.94 (0.73, 1.25)[ab] | 0.91 (0.71, 1.17)[a] | 0.81 (0.67, 1.00) |
| Highest / reference creatinine, mean (SD) | 1.22 (0.61) | 1.91 (1.32) | 2.29 (1.83)[abc] | 2.14 (1.47)[ab] | 1.65 (0.76)[a] | 1.09 (0.14) |
| Count of nephrotoxic drugs, mean (SD) | | | | | | |
| Within 2 days after hospital admission | 0.55 (0.83) | 0.71 (0.93)[a] | 0.74 (0.94)[abc] | 0.78 (0.97)[ab] | 0.67 (0.91)[a] | 0.52 (0.81) |
| Within 3 days after hospital admission | 0.60 (0.87) | 0.78 (0.98)[a] | 0.81 (0.99)[abc] | 0.88 (1.04)[ab] | 0.74 (0.96)[a] | 0.57 (0.85) |
| Between hospital admission and first AKI onset | 0.70 (1.00) | 0.70 (1.00) | 0.74 (1.03)[abc] | 0.79 (1.07)[b] | 0.66 (0.97) | NA |

Abbreviations. SD, standard deviation; IQR, interquartile range; RRT, renal replacement therapy; AKI, Acute Kidney Injury.

[a] p<0.05 compared to no AKI
[b] p<0.05 compared to Rapidly reversed AKI
[c] p<0.05 compared to Persistent AKI with renal recovery
[d] Cardiovascular disease was considered if there was a history of congestive heart failure, coronary artery disease of peripheral vascular disease.

female patients was lower for the ICU cohort (44% vs 54%) with a higher proportion of African Americans represented (25% vs 20%) (Supplemental Tables 2 and 3).

***CKD and AKI incidence and characteristics***

Of 2,187,538 analyzed total patient encounters, 14% (n = 301,464) had AKI during hospitalization. Of this sample, 63% (n = 190,244), 21% (63,519), and 16% (n = 47,701) reached a maximum of stage 1, stage 2, and stage 3, respectively (Supplemental Table 4). Among those with AKI, 69% (n = 163,108) had rapidly reversed AKI while the remaining 31% (138,356) had persistent AKI (Table 1). Twenty two percent of the study cohort (n = 475,643) had pre-existing CKD. The majority of CKD diagnoses were identified based on a patient's medical history (79%), while 18% of CKD diagnoses were based on creatinine criteria and the remaining 2% were based on post-kidney transplant status, respectively (Supplemental Table 5). The most common CKD stage was G2 (31%) while the least common stage was G5 (2%). For our entire cohort, patients with AKI were almost twice as likely to have pre-existing CKD compared to patients without AKI (Supplemental Tables 1-3).

***Inpatient clinical characteristics***

Stage 1 AKI was the most common stage for patients with persistent AKI without recovery (41%), for persistent AKI with recovery (46%) and for those with rapidly reversed AKI (80%). This trend highlights a higher prevalence of severe AKI stages within the persistent AKI group, particularly among those without recovery. (Table 2).

**Table 2.** Renal characteristics, resource utilization, and hospital outcomes during entire hospitalization by trajectories of AKI in all cohort.

| Variables | All subjects (N= 2,187,254) | AKI (N = 301,464, %14) | Persistent AKI without renal recovery (N= 92,607, %4) | Persistent AKI with renal recovery (N=45,749, %2) | Rapidly reversed AKI (N=163,108, %8) | No AKI (N=18,85 790, %86) |
|---|---|---|---|---|---|---|
| **Renal characteristics during entire hospitalization** | | | | | | |
| **Worst AKI Staging, n (%)** | | | | | | |
| Stage 1 | 190,244 (9) | 190,244 (63)[a] | 38,182 (41)[abc] | 20,979 (46)[ab] | 131083 (80)[a] | 0 (0) |
| Stage 2 | 63,519 (3) | 63,519 (21)[a] | 27,944 (30)[abc] | 13,226 (29)[ab] | 22,349 (14)[a] | 0 (0) |
| Stage 3 | | | | | | |
| Stage 3 without RRT | 45,707 (2) | 45,707 (15)[a] | 24,935 (27)[abc] | 11,124 (24)[ab] | 9,648 (6)[a] | 0 (0) |
| Stage 3 with RRT | 1,994 (0) | 1,994 (1)[a] | 1,546 (2)[abc] | 420 (1)[ab] | 28 (0)[a] | 0 (0) |
| AKI duration, days, median (IQR) | 0 (0, 0) | 2 (1, 4)[a] | 5 (3, 8)[b] | 4 (3, 6)[ab] | 1 (1, 2) | 0 (0, 0) |
| Recurrent AKI, n (%) | 23,938 (1) | 23,938 (8)[a] | 9,459 (10)[abc] | 7,529 (16)[ab] | 6,950 (4)[a] | 0 (0) |
| No renal recovery at discharge/death, n (%) | 135,640 (6) | 127,765 (42)[a] | 92,607 (100)[abc] | 995 (2)[ab] | 34,163 (21)[a] | 7,875 (0) |
| **Resource utilization during entire hospitalization** | | | | | | |
| Hospital days, median (IQR) | 3 (1, 5) | 5 (3, 10)[a] | 5 (3, 10)[abc] | 10 (6, 18)[ab] | 4 (2, 8)[a] | 3 (1, 5) |
| Admission to ICU, n (%) | 83,478 (4) | 24,477 (8)[a] | 8,099 (9)[abc] | 5,222 (11)[ab] | 11,156 (7) | 59,001 (3) |
| Mechanical Ventilation, n (%) | 84,466 (4) | 38,020 (13)[a] | 15,731 (17)[abc] | 9,308 (20)[ab] | 12,981 (8)[a] | 46,446 (2) |
| Vasopressor or inotropes used, n (%) | 126,199 (6) | 31,894 (11)[a] | 11,660 (13)[abc] | 6,876 (15)[ab] | 13,358 (8)[a] | 94,305 (5) |
| **Hospital disposition, n (%)** | | | | | | |

| Variables | All subjects (N= 2,187,254) | AKI (N = 301,464, %14) | Persistent AKI without renal recovery (N= 92,607, %4) | Persistent AKI with renal recovery (N=45,749, %2) | Rapidly reversed AKI (N=163,108, %8) | No AKI (N=18,85790, %86) |
|---|---|---|---|---|---|---|
| Hospital mortality | 19,541 (1) | 11,666 (4)[a] | 9,204 (10)[abc] | 995 (2)[ab] | 1,467 (1)[a] | 7,875 (0) |
| Another hospital, LTAC, SNF, Hospice | 221,561 (10) | 63,526 (21)[a] | 26,620 (29)[ab] | 13,023 (28)[ab] | 23,883 (15)[a] | 158,035 (8) |
| Home/Rehab | 1,946,152 (89) | 226,272 (75)[a] | 56,783 (61)[abc] | 31,731 (69)[ab] | 137,758 (84)[a] | 1,719,880 (91) |
| 30-day outcomes (among survivors), n (%) | | | | | | |
| Death in 30 days of discharge | 7,007 (0) | 2,083 (1)[a] | 773 (1)[ab] | 358 (1)[ab] | 952 (1)[a] | 4,924 (0) |
| Trajectory group for encounter with readmission within 30 days of discharge | | | | | | |
| Persistent AKI with no renal recovery | 14,791 (3) | 14,648 (17)[a] | 10,654 (48)[abc] | 1,083 (7)[ab] | 2,911 (6)[a] | 10,998 (2) |
| Persistent AKI with renal recovery | 48,595 (9) | 8,572 (10)[a] | 1,110 (5)[abc] | 5,637 (37)[ab] | 1,825 (4)[a] | 6,219 (1) |
| Rapidly reversed AKI | 471,134 (84) | 27,580 (33)[a] | 3,267 (15)[abc] | 1,922 (13)[ab] | 22,391 (48)[a] | 21,015 (4) |
| No AKI | 7,007 (0) | 33,594 (40)[a] | 7,240 (33)[abc] | 6,477 (43)[a] | 19,877 (42)[a] | 437,540 (92) |
| **Other complications during entire hospitalization** | | | | | | |
| Venous Thromboembolism, n (%) | 82,574 (4) | 19,840 (7)[a] | 6,189 (7)[abc] | 5,005 (11)[ab] | 8,646 (5)[a] | 62,734 (3) |
| Sepsis, n (%) | 153,014 (7) | 56,295 (19)[a] | 19,866 (21)[abc] | 13,055 (29)[ab] | 23,374 (14)[a] | 96,719 (5) |
| Cardiovascular complication, n (%) | 133,967 (6) | 48,952 (16)[a] | 16,782 (18)[abc] | 9,907 (22)[ab] | 22,263 (14)[a] | 85,015 (5) |

| Variables | All subjects (N= 2,187,254) | AKI (N = 301,464, %14) | Persistent AKI without renal recovery (N= 92,607, %4) | Persistent AKI with renal recovery (N=45,749, %2) | Rapidly reversed AKI (N=163,108, %8) | No AKI (N=18,85790, %86) |
|---|---|---|---|---|---|---|
| Thirty-day mortality, n (%) | 25,641 (1) | 13,052 (4)[a] | 9,523 (10)[abc] | 1,201 (3)[ab] | 2,328 (1)[a] | 12,589 (1) |
| One-year mortality, n (%) | 66,692 (3) | 23,914 (8)[a] | 12,476 (13)[abc] | 3,672 (8)[ab] | 7,766 (5)[a] | 42,778 (2) |
| Three-year mortality, n (%) | 96,388 (4) | 30,375 (10)[a] | 13,865 (15)[abc] | 4,861 (11)[ab] | 11,649 (7)[a] | 66,013 (4) |

Abbreviations. SD, standard deviation; IQR, interquartile range; RRT, renal replacement therapy; ICU, intensive care unit; LTAC, long-term acute care hospital; SNF, skilled nursing facility; AKI, Acute Kidney Injury; NA, not applicable.

[a] $p<0.05$ compared to no AKI

[b] $p<0.05$ compared to Rapidly reversed AKI

[c] $p<0.05$ compared to Persistent AKI with renal recovery

Compared to patients with no AKI, subjects with AKI had longer hospital stays, higher rate of ICU admissions, longer duration on mechanical ventilation, increased inotrope use, higher hospital mortality rates, and a reduced likelihood of being discharged home post-hospitalization. Additionally, those with AKI were more likely to be older and have additional comorbidities, such as hypertension, cardiopulmonary disease, diabetes, or pre-existing kidney disease. Thromboembolism, sepsis, and cardiovascular complications were more frequent in AKI patients where the highest prevalence for these conditions were observed among persistent AKI patients with recovery. Additionally, within 30 days of discharge, 60% of patients with AKI during hospitalization had recurrent AKI.

In addition, AKI patients were also significantly more likely to have other comorbidities. Within the first 48 hours of their hospital presentation, persistent AKI patients had greater blood urea nitrogen (mean range 35–40 mg/dL, SD range 26–27 mg/dL), serum creatinine (median range 1.51–1.80 mg/dL, interquartile range [IQR] 1.06–2.75 mg/dL), and serum creatinine to reference creatinine ratios (mean range 2.14–2.29, SD range 1.47–1.83). Similarly, persistent AKI patients had higher mean serum creatinine values (mean range 2.56–3.05, SD range 1.84–2.31) than those with rapidly reversed AKI within two days following onset (Supplemental Table 6). Overall, persistent AKI patients were exposed to nephrotoxic drugs more frequently, compared to patients with rapidly reversed AKI within the second and third days of their hospital presentation, including the period between admission and first AKI incident.

### ***Resource utilization and hospital and long-term mortality***

The median length of hospital stay for all patients was 3 days (IQR 1–5). In this cohort, persistent AKI with recovery group had a longer length of stay (10 days), as well as a higher proportion of mechanical ventilation utilization (20%), ICU admission (10%) and vasopressor or inotrope administration (15%) than those with rapidly reversed AKI. Mortality rates were highest for persistent AKI without recovery group, with 10% for hospital mortality, 10% for 30-day mortality, 13% for 1-year mortality, and 15% for 3-year mortality. Meanwhile, the persistent AKI with recovery group had the second highest mortality rates, with 2% for hospital mortality, 3% for 30-day mortality, 8% for 1-year mortality, and 11% for 3-year mortality (Table 2). Readmission and new CKD onset proportions within 90 days and 1 year following discharge were higher for patients with persistent AKI (Supplemental Tables 7–9). In ICU cohort, patients with AKI had longer a length of stay (9 days [IQR 4–17]) compared to those without AKI (5 days [IQR 2–9]) with increased mechanical ventilation utilization and mortality during hospital and over 30-day, 1-year, and 3-year following discharge (Supplemental Table 10). Complications during hospitalization, 30-day, 1-year and 3-year mortality was the most common and hospital length of stay was the highest for patients with persistent AKI in ICU cohort. Despite reduced prevalence in complication and mortality variables along with shorter length of stay, similar pattern remained for patients with persistent AKI in non-ICU group (Supplemental Table 11).

Regardless of being admitted to the ICU, odds of death during hospitalization for persistent AKI patients were significantly greater compared to the no AKI group (Supplemental Table 12). Specifically, compared to no AKI group, odds ratios (OR) for patients with persistent AKI without recovery ranged between 16.02 and 26.31 while OR range was 3.11–5.30 for persistent AKI patients with recovery after adjusting for sex, age, CCI and AKI severity. One-year mortality among persistent AKI patients (%12, n = 16,148) was more than two-folds and five-folds the one-year mortality rate for rapidly reversed AKI patients (5%, n = 7,766) and patients without AKI (2%, n=42,778), respectively. Majority of the admitted patients had early AKI and of those with early AKI, more than half of the trajectories had renal recovery (Figure 1A). Survival rates post-discharge were the lowest for persistent AKI without recovery, while persistent AKI with recovery had the second worst and patients without AKI had the best survival outcome (Figure 1B).

With patients with no AKI as the reference group, hazard ratios for persistent AKI without recovery were 8.5 (95% CI 8.3–8.7) and 3.7 (95% CI 3.6–3.8) in unadjusted and adjusted models for the overall cohort, respectively (Supplemental Table 13). In the ICU cohort, one-year survival was notably lower for persistent AKI patients without renal recovery. All-cause mortality rates in adjusted models were more than four-folds and three-folds for ICU and non-ICU cohorts, respectively. Hazard ratios were slightly elevated when the models adjusted based upon AKI severity.

We further elaborated AKI trajectories by including AKI severity in addition to renal recovery and duration details, which resulted in seven distinct trajectory groups. One-year survival for AKI group was significantly reduced compared to patients with no AKI (Figure 2). Similarly, severe AKI was associated with reduced survival in comparison to mild and no AKI. Survival was the lowest for severe, persistent AKI without recovery (76%), followed mild, persistent AKI without recovery (84%).

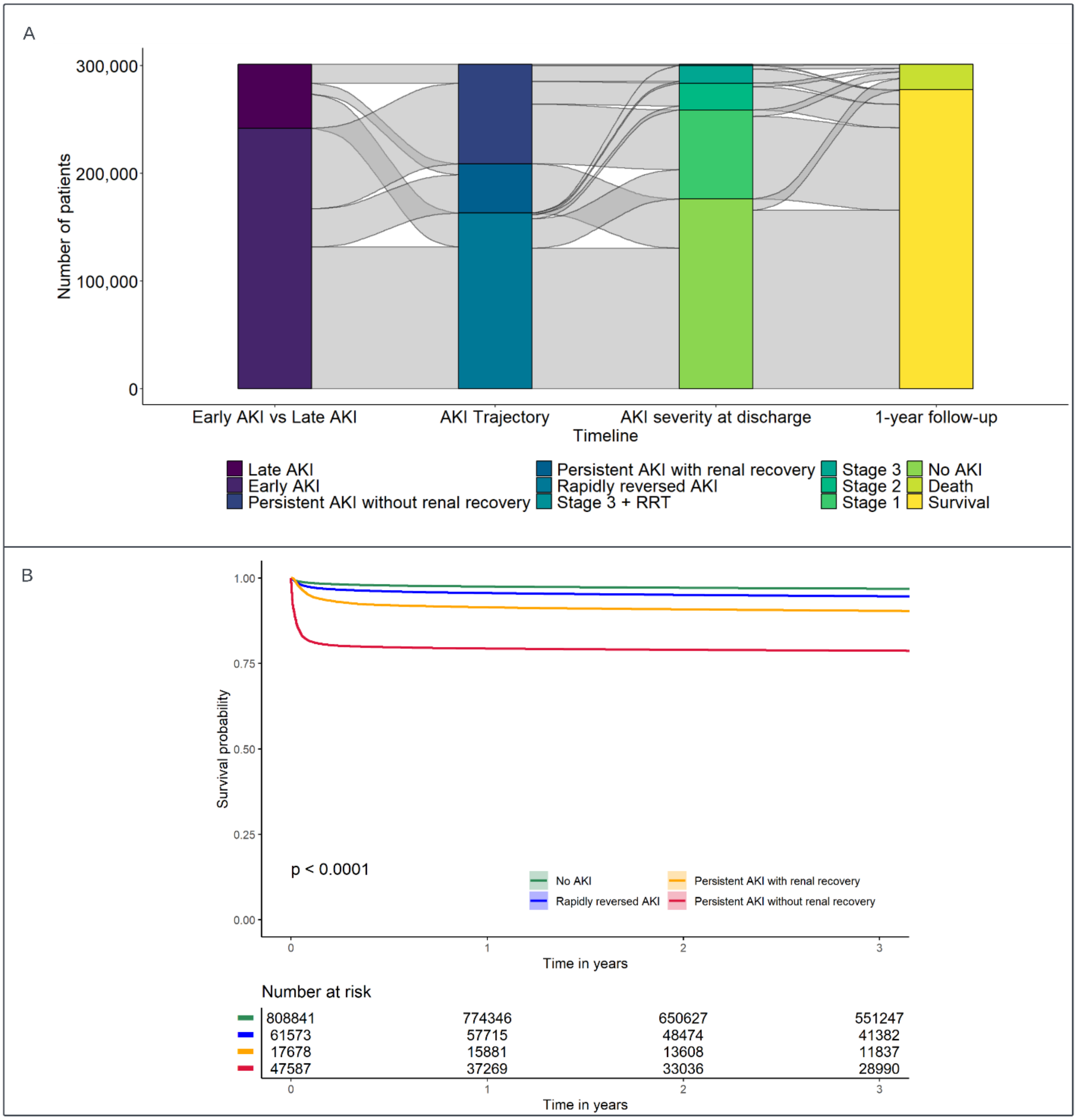

**Figure 1. Patient outcomes during hospital admission and up to 3 years after discharge among hospitalized adult patients. A.** Acute kidney injury (AKI) trajectories outcomes 1-year following discharge for all hospitalized adult patients. **B.** Adjusted Kaplan-Meier survival curves and number at risk for each AKI trajectory group. Computed results were adjusted for age, sex, race, Charlson Comorbidity Index score, and receiving mechanical ventilation treatment for ≥ 2 days and being admitted to intensive care unit for ≥ 2 days.

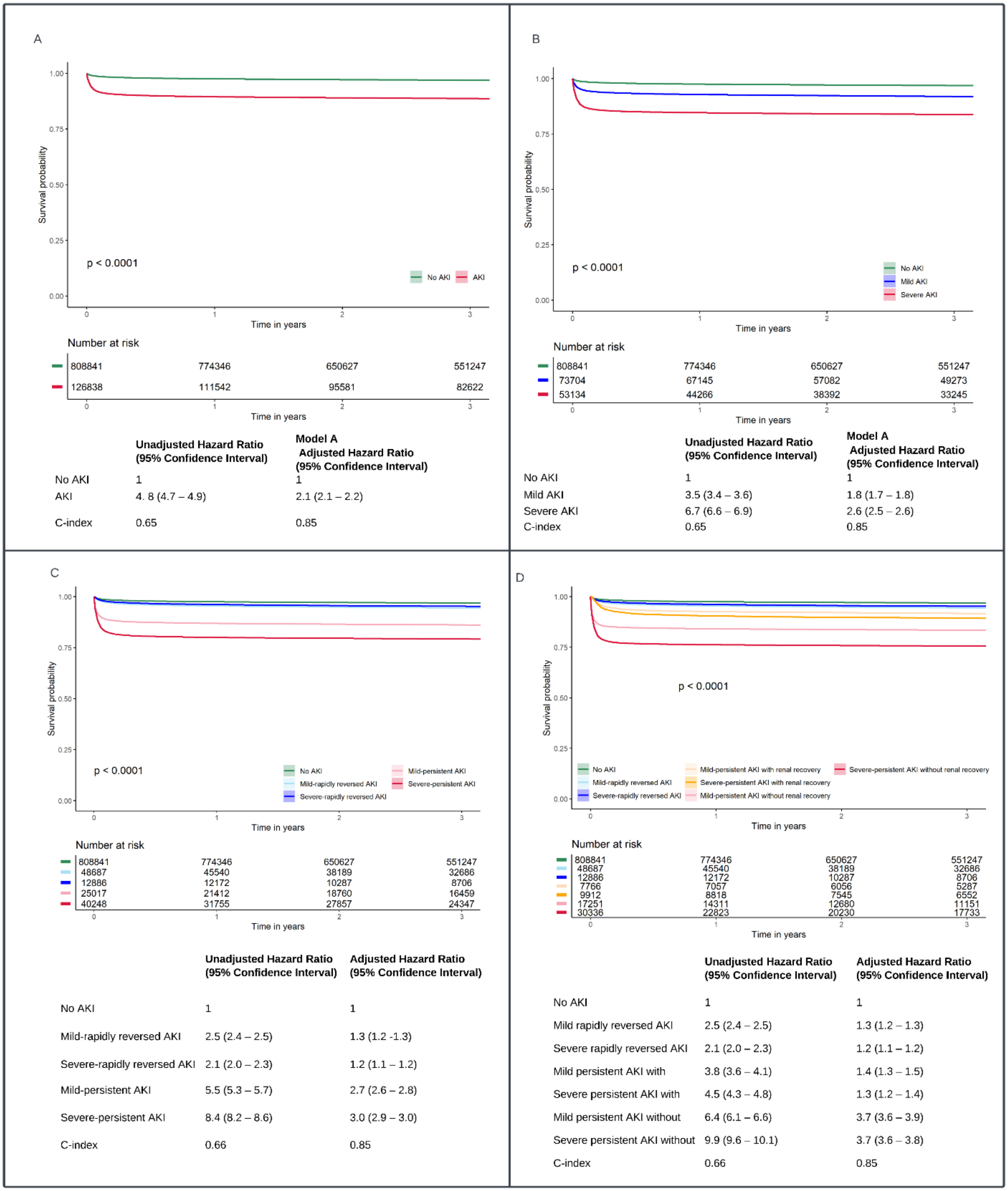

**Figure 2. Adjusted Kaplan-Meier survival curves and number at risk by acute kidney injury (AKI) subphenotypes. A.** Subphenotypes stratified by no AKI vs any AKI. **B.** Subphenotypes stratified by AKI severity. **C.** Subphenotypes stratified by AKI

severity and duration. **D.** Subphenotypes stratified by stratified by AKI severity and trajectories, indicated by AKI duration and recovery from AKI.

Difference in survival rates was not significant for rapidly reversed AKI group and persistent AKI with recovery group, however survival rates were significantly less for mild-persistent AKI without recovery and severe-persistent AKI without recovery trajectories. In addition, mild persistent AKI was associated with significantly poorer outcomes compared to severe rapidly reversed AKI. While persistent AKI without renal recovery had the lowest survival for all cohort as well as ICU and non-ICU subgroups, difference was more substantial in the ICU subgroup. (Figure 3, Supplemental Figure 2 and 3)

## Discussion

In this study, we presented a thorough landscape of short- and long-term outcomes associated with AKI trajectories in hospitalized patients by compiling a comprehensive taxonomy developed around severity, duration and recovery characteristics. We further offered a layer of granularity to epidemiologic characteristics of AKI trajectories by performing analyses that were extended to ICU and non-ICU subgroups. Overall, 14% of this cohort of hospitalized patients had AKI during their admission, of which 31% had persistent AKI. The proportion of AKI patients remained similar for those who have never been in the ICU (13%) while substantially increasing for the ICU subgroup (29%). Moreover, the percentage of persistent AKI was calculated as 45% and 59% for non-ICU and ICU subgroups, respectively.

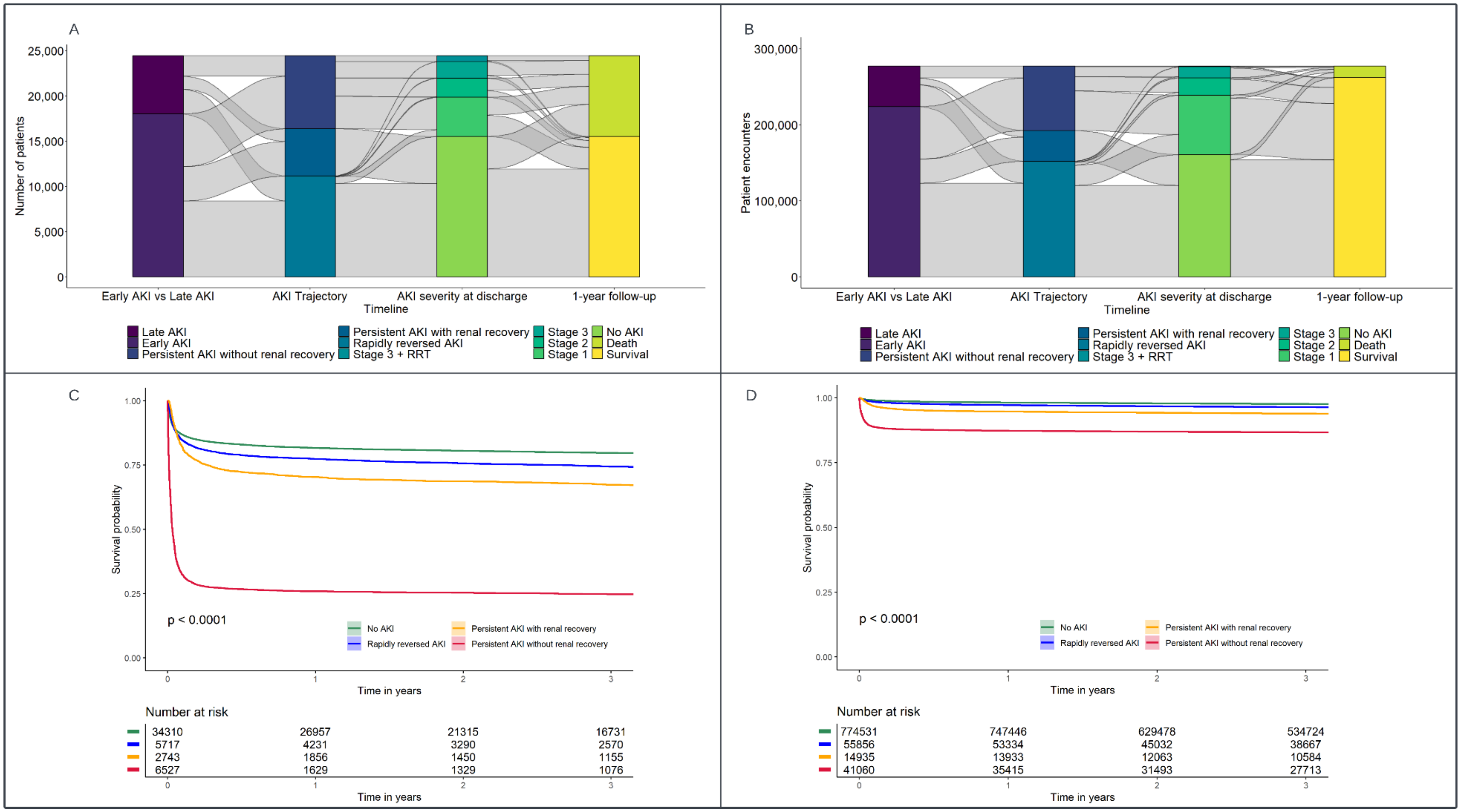

**Figure 3. Patient outcomes during hospital admission and up to 3 years after discharge among hospitalized adult patients stratified by intensive care unit (ICU) admission during their stay. A.** Acute kidney injury (AKI) trajectories and outcomes 1-year after discharge for all patients who have been admitted to the ICU during their hospital stay. **B.** Acute kidney injury (AKI) trajectories and outcomes 1-year after discharge for all patients who have not been admitted to the ICU during their hospital stay. **C.** Adjusted Kaplan-Meier curves by AKI trajectories and number at risk for patients who

have been admitted to the ICU during their hospital stay. **D.** Adjusted Kaplan-Meier curves by AKI trajectories and number at risk for patients who have not been admitted to the ICU during their hospital stay.

Sepsis, cardiac complications and venous thromboembolism were more frequent in persistent AKI patients for the overall cohort and its separate non-ICU and ICU subgroups. We observed deteriorated one-year mortality rates for poorer AKI trajectories, as the one-year mortality rate was the lowest for rapidly reversed AKI patients (5%), which was followed by persistent AKI with rapid recovery (8%) and persistent AKI without recovery (13%). Patients with persistent AKI trajectories were admitted to ICU, were administered vasopressor and received mechanical ventilation more frequently compared to those with rapidly reversed AKI. Regardless of recovery and severity, subjects with persistent AKI had the worst one-year and three-year survival rates, where the worst outcomes were observed in the severe persistent AKI without recovery group.

Research focusing on different aspects in describing AKI trajectories had been done for hospitalized patients before. Specifically, relationships between patient outcomes and AKI duration were investigated in several studies.[26] A multicenter cohort study for 1,538 subjects, was conducted to assess long-term outcomes for non-resolving and resolving AKI classifications and indicated worse mortality, CKD progression and long-term dialysis outcomes for patients who had non-resolving AKI.[27] Another study performed for a cohort of 2,143 critical care patients reported increase in mortality risk with AKI duration.[28] A study conducted for a cohort of 16,986 ICU patients found that persisting AKI was associated with worse mortality.[29] Similarly, AKI duration was shown to be proportional to mortality in a surgical cohort of 4,987 patients.[30] That observation was relevant for a cohort of 123 AKI patients in a cohort of 10,250 hospitalized subjects.[31] Despite their important merits, those studies were previously criticized for being exclusive to non-ICU or ICU patients.[16] A recent multicenter, large cohort study investigated associations between patient outcomes with persistent AKI for those who had Stage 2 and 3, and performed analyses for all cohort, ICU and non-ICU subgroups.[16] According to their findings, persistent AKI was associated with worse outcomes occurred during hospitalization and 30-days following discharge for a cohort of 126,528 patients with Stage 2 and Stage 3 AKI. Koyner et al. (2022) presented one of the closest works to our current study, however, reported inferences on survival in their study were limited by observations obtained during hospitalization and 30-days after discharge. We performed similar analyses for all patients admitted to a single center, in addition to ICU and non-ICU subgroups, and obtained the worst long-term survival, i.e. 3-years, for persistent AKI without recovery trajectory.[8] In addition to a cohort with substantially larger number of subjects, the multicenter study design of this current work is the first and foremost difference from the one we presented before in Ozrazgat-Baslanti et al. (2021).

Our motivation to perform this work was multifactorial. Given the findings of the recent retrospective cohort study from University of Florida Health (UFH) Integrated Data Repository, which provided evidence for correlations between physiological indicators with AKI trajectories and overall patient mortality, similar results obtained by performing analyses for patient data from large multicenter OneFlorida Data Trust would potentially support that these findings are more generalizable. To the best of our knowledge, this is the first work that considered AKI severity, persistence and renal recovery characteristics simultaneously in a multicenter and large cohort setting with a long-term follow-up. Specifically, patients were geographically dispersed across urban and rural areas in the OneFlorida Data Trust, which supports the generalizability of the findings presented in this study. Another rationale for performing this study was to demonstrate the usability of previously validated AKI phenotyping algorithm for a different common data model that is PCORnet, as the algorithm operates on detailed reference creatinine assignment and CKD identification procedures in line with KDIGO and ADQI guidelines[14, 32]. Standardized and systematic AKI phenotyping algorithms that are able to operate on different common data models would contribute to more compatible and transferable treatment strategies.

One important limitation of this study is that AKI identification relies on serum creatinine criteria rather than hourly urine output, since urine output measurements were not collected as consistently and reliable as in wards as usually performed in the critical care setting. Another limitation of this study could be that relying on ICD-9, ICD-10 and CPT codes when identifying ICU admission and mechanical ventilation utilization indicators as the exact timings of in and out of ICU along with mechanical ventilation were not available. Since these codes were primarily collected for billing purposes, relying on those codes did not allow us to perform more granular ICU and mechanical ventilation analyses. Finally, analyzed OneFlorida+ data was obtained from patients mostly residing in state of Florida which may affect generalizability of the findings presented in this work to the populations from different areas.

## Conclusions

Our large multicenter cohort based retrospective analyses demonstrated the association of persistent AKI with worse clinical outcomes, including reduced survival rates, increased resource utilization, and increased CKD progression even after adjusting for severity. Our findings underline the need for strategies that optimize medical care to prevent persistent AKI and improve patient outcomes.

## Data availability

The data used in this study were obtained from the OneFlorida+ Clinical Research Network (https://onefloridaconsortium.org/). Researchers may access the data with an approved study protocol and data use agreement (DUA) at https://onefloridaconsortium.org/front-door/prep-to-research-data-query/.

# Supplemental Material

Adiyeke E., Ren Y., Fogel S., Rashidi P., Segal M., Shenkman E., Bihorac A., Ozrazgat-Baslanti T. **Epidemiology, Trajectories, and Outcomes of Acute Kidney Injury Among Hospitalized Patients: A Retrospective Multicenter Large Cohort Study**

This supplemental material has been provided by the authors to give readers additional information about their work.

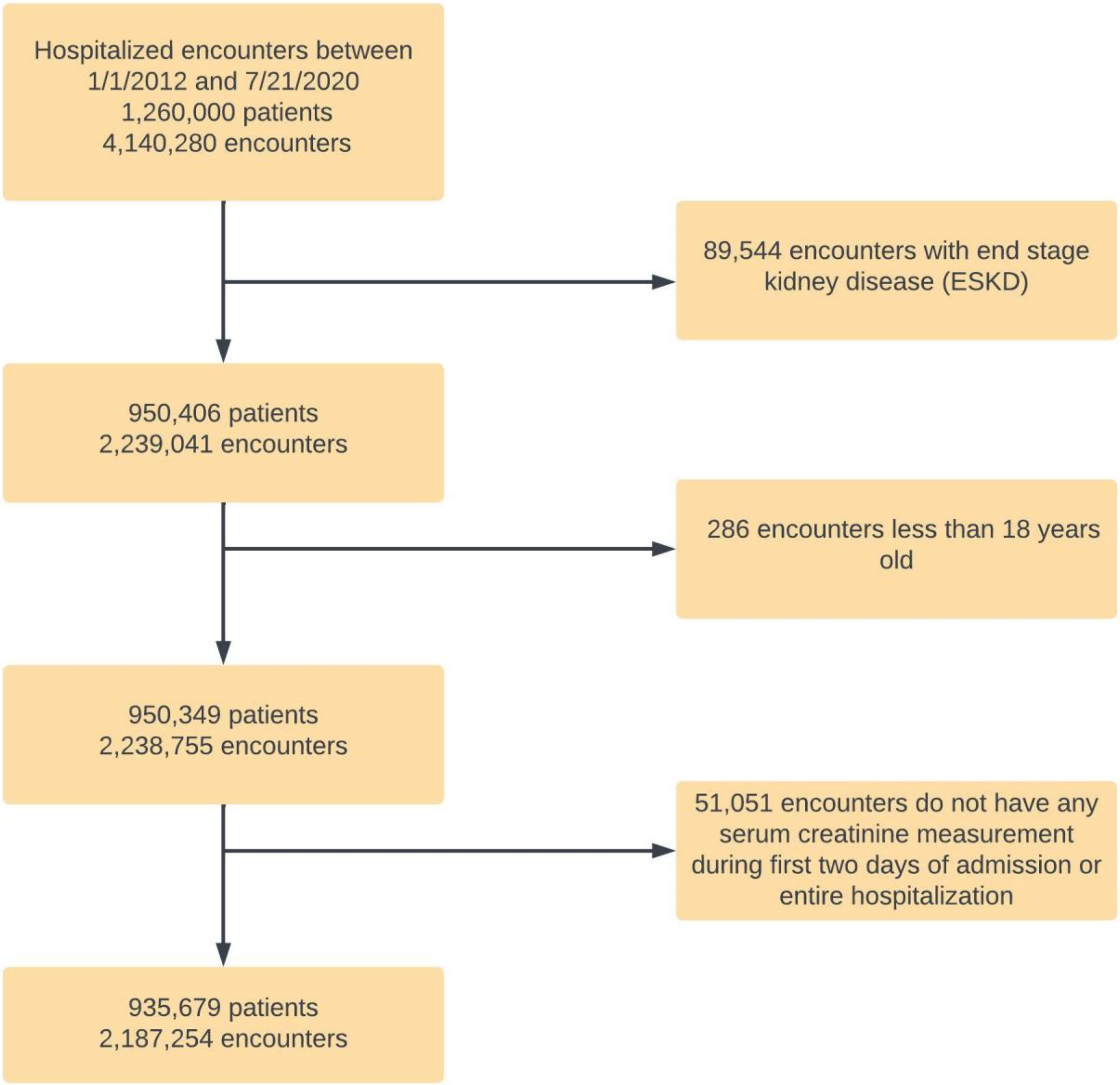

**Supplemental Figure 1.** Cohort derivation.

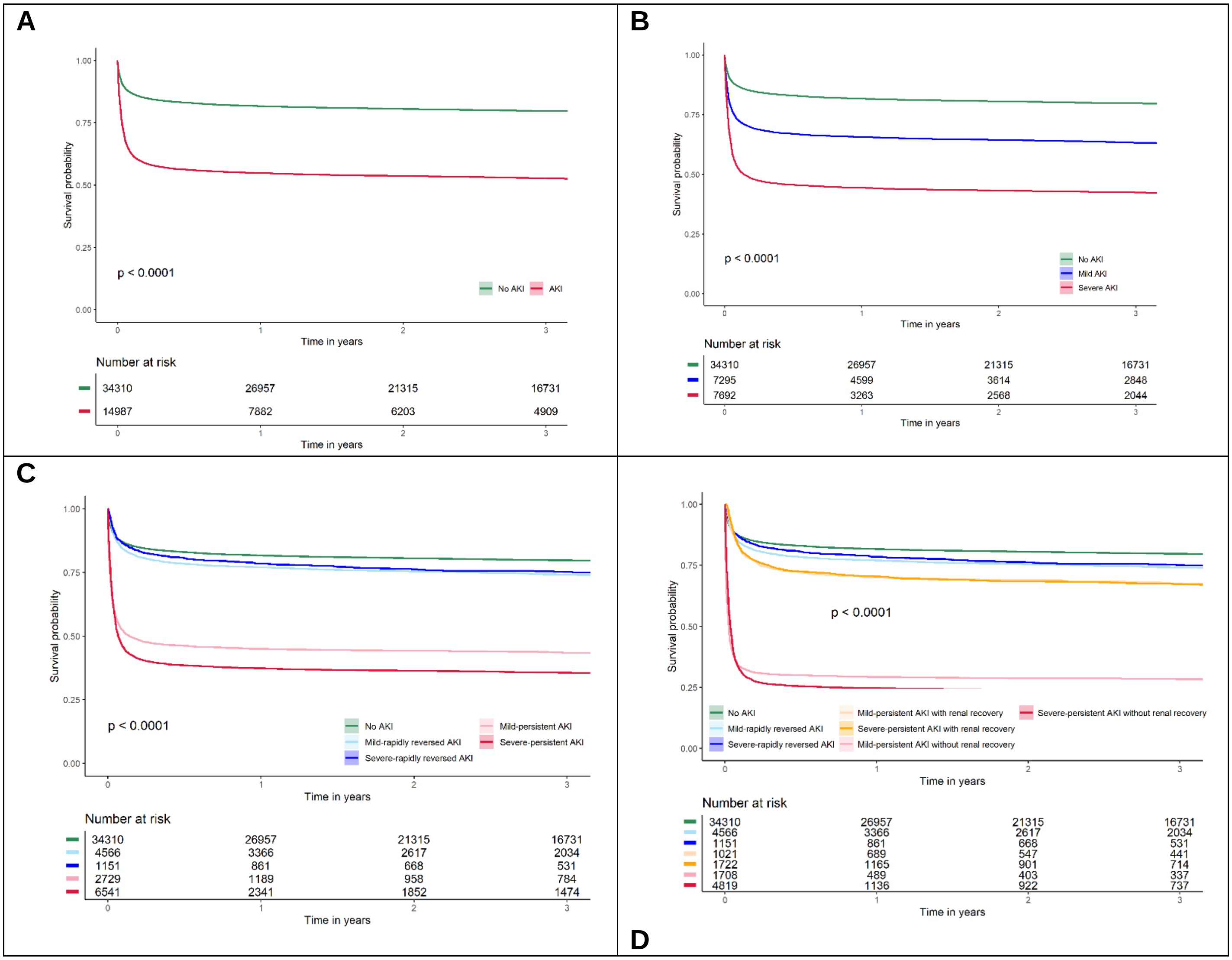

**Supplemental Figure 2.** Adjusted Kaplan-Meier survival curves and number at risk by AKI sub-phenotypes, in patients who have been admitted to ICU during hospitalization, obtained stratifying by

A. No AKI vs. Any AKI
B. AKI stratified by severity
C. AKI stratified by severity and duration
D. AKI stratified by severity and trajectories of AKI using duration and recovery of AKI

Propensity score based inverse weighting was used to plot adjusted Kaplan Meier curves where propensity of being in a trajectory group was calculated using multinomial logistic model that included age, gender, ethnicity, and Charlson comorbidity score. Adjusted hazard ratios were obtained adjusting for the same variables as well as need for mechanical ventilation and need for intensive care unit admission.

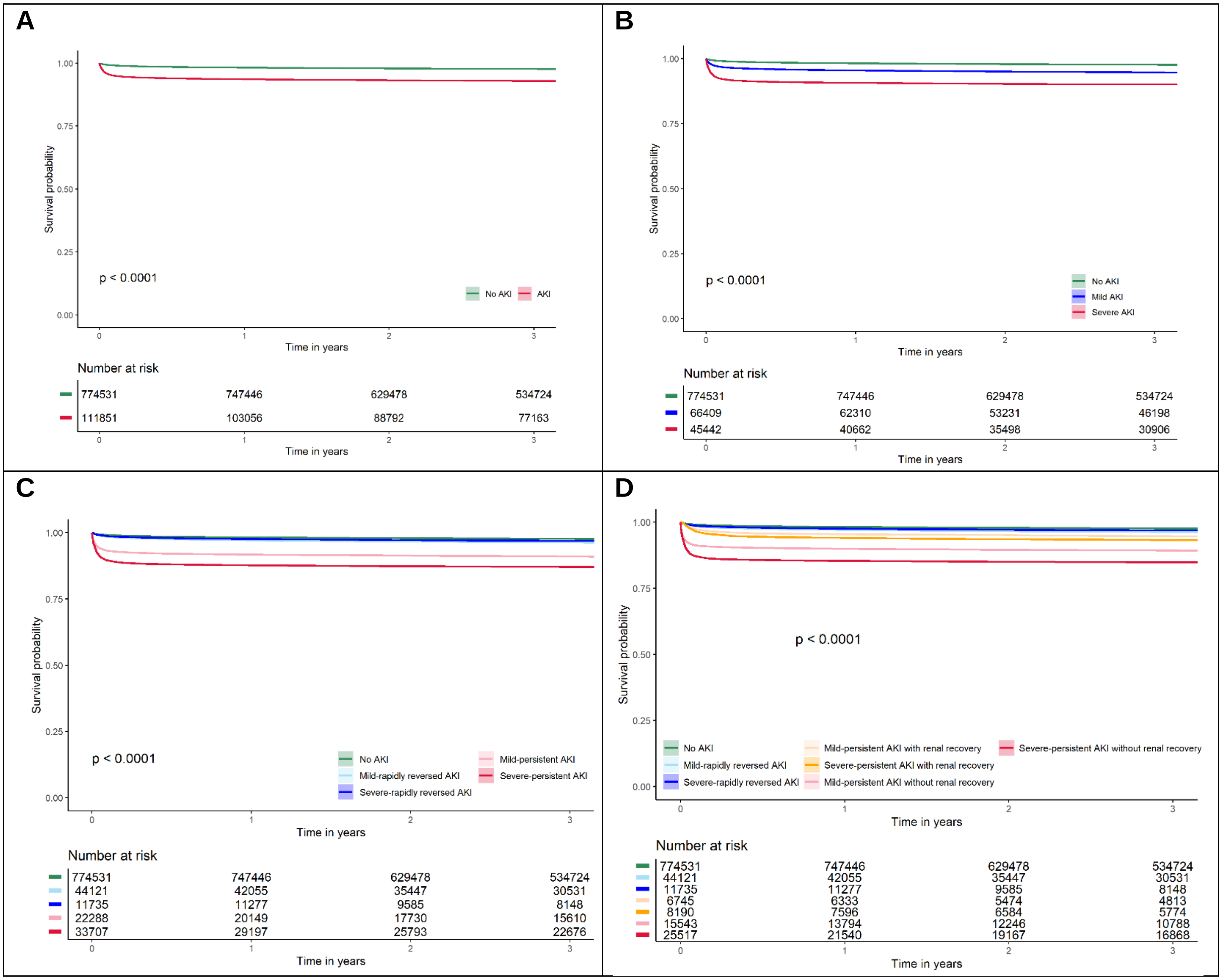

**Supplemental Figure 3.** Adjusted Kaplan-Meier survival curves and number at risk by AKI sub-phenotypes, in patients who have not been admitted to ICU during hospitalization, obtained stratifying by

A. No AKI vs. Any AKI

B. AKI stratified by severity

C. AKI stratified by severity and duration

D. AKI stratified by severity and trajectories of AKI using duration and recovery of AKI

Propensity score based inverse weighting was used to plot adjusted Kaplan Meier curves where propensity of being in a trajectory group was calculated using multinomial logistic model that included age, gender, ethnicity, and Charlson comorbidity score. Adjusted hazard ratios were obtained adjusting for the same variables.

**Supplemental Table 1.** Baseline characteristics by trajectory groups in all cohort.

| Variables | All Cohort (N=2,187,254) | AKI (N= 301,464, %14) | Persistent AKI without renal recovery (N=92,607, %4) | Persistent AKI with renal recovery (N=45,749, %2) | Rapidly Reversed AKI (N=163,108, %8) | No AKI (N= 1,885,790, %86) |
|---|---|---|---|---|---|---|
| **Preadmission Clinical Characteristics** | | | | | | |
| Age, mean (SD), years | 59 (22) | 69 (24)[a] | 74 (27)[abc] | 68 (22)[ab] | 67 (23)[a] | 58 (22) |
| Female sex, n (%) | 1,167,105 (53) | 157,121 (52)[a] | 51,895 (56)[abc] | 23,035 (50)[a] | 82,191 (50)[a] | 1,009,984 (54) |
| Race, n (%) | | | | | | |
| African American | 432,591 (20) | 65,225 (22)[a] | 19,048 (21)[abc] | 10,133 (22)[a] | 36,044 (22)[a] | 367,366 (19) |
| White | 1,476,166 (67) | 199,795 (66)[a] | 61,919 (67)[abc] | 30,000 (66)[a] | 107,876 (66)[a] | 1,276,371 (68) |
| Primary Insurance, n (%) | | | | | | |
| Medicare | 400,384 (18) | 78,673 (26)[a] | 24,934 (27)[abc] | 11,999 (26)[ab] | 41,740 (26)[a] | 321,711 (17) |
| Private | 709,566 (32) | 82,008 (27)[a] | 23,645 (26)[ab] | 11,842 (26)[ab] | 46,521 (29)[a] | 627,558 (33) |
| Medicaid | 20,0985 (9) | 29,732 (10)[a] | 8,477 (9)[bc] | 4,815 (11)[ab] | 16,440 (10)[a] | 171,253 (9) |
| Uninsured | 70,883 (3) | 6,079 (2)[a] | 1,558 (2)[ab] | 699 (2)[ab] | 3,822 (2)[a] | 64,804 (3) |
| Other | 732,199 (33) | 98,702 (33)[a] | 32,707 (35)[abc] | 156,86 (34)[ab] | 50,309 (31)[a] | 633,497 (34) |
| Charity | 73,234 (3) | 6,270 (2)[a] | 1,286 (1)[ab] | 708 (2)[ab] | 4,276 (3 [a]) | 66,964 (4) |
| **Comorbidities, n (%)** | | | | | | |
| Hypertension | 362,545 (17) | 89,852 (30)[a] | 24,840 (27)[abc] | 16,053 (35)[ab] | 48,959 (30)[a] | 272,693 (14) |
| Chronic pulmonary disease | 604,472 (28) | 96,967 (32)[a] | 26,740 (29)[abc] | 15,760 (34)[ab] | 54,467 (33)[a] | 507,505 (27) |
| Cardiovascular disease * | 562,903 (26) | 115,883 (38)[a] | 32,766 (35)[abc] | 19,990 (44)[ab] | 63,127 (39)[a] | 447,020 (24) |
| Diabetes mellitus | 599,788 (27) | 113,184 (38)[a] | 30,730 (33)[abc] | 18,134 (40)[a] | 64,320 (39)[a] | 486,604 (26) |
| Chronic kidney disease | 475,640 (22) | 111,501 (37)[a] | 27,982 (30)[abc] | 21,240 (46)[ab] | 62,279 (38)[a] | 364,139 (19) |
| Moderate/Severe (≥ G-Stage 3) | 263,644 (55) | 64,829 (58)[a] | 15,022 (54)[abc] | 12,716 (60)[a] | 37,091 (60)[a] | 198,815 (55) |

| **Variables** | **All Cohort (N=2,187,254)** | **AKI (N= 301,464, %14)** | **Persistent AKI without renal recovery (N=92,607, %4)** | **Persistent AKI with renal recovery (N=45,749, %2)** | **Rapidly Reversed AKI (N=163,108, %8)** | **No AKI (N= 1,885,790, %86)** |
|---|---|---|---|---|---|---|
| Preadmission estimated glomerular filtration rate (mL/min per 1.73 $m^2$), median (IQR) | 56.59 (40.87, 74.76) | 54.24 (37.75, 74.35)[a] | 57.03 (38.45, 81.69)[bc] | 53.04 (36.75, 72.74)[ab] | 53.71 (37.88, 72.30)[a] | 57.24 (41.76, 74.87) |
| Transfer from another hospital | 172,607 (8) | 34,705 (12)[a] | 12,792 (14)[ab] | 6,618 (14)[ab] | 15,295 (9)[a] | 137,902 (7) |
| Surgery on admission day | 43,421 (2) | 6,141 (2)[a] | 1,532 (2)[bc] | 1,071 (2)[ab] | 3,538 (2)[a] | 37,280 (2) |
| Surgery at any time | 124,813 (6) | 18,860 (6)[a] | 4,811 (5)[bc] | 4,155 (9)[ab] | 9,894 (6)[a] | 105,953 (6) |
| **Location/Type of surgery**, n (%)** | | | | | | |
| Musculoskeletal System | 30,117 (24) | 2,975 (16)[a] | 732 (15)[ab] | 551 (13)[ab] | 1,692 (17)[a] | 27,142 (26) |
| Integumentary System | 12,253 (10) | 1,498 (8)[a] | 430 (9)[ac] | 268 (6)[a] | 800 (8)[a] | 10,755 (10) |
| Digestive System | 25,685 (21) | 3,707 (20)[a] | 864 (18)[ab] | 834 (20) | 2,009 (20) | 21,978 (21) |
| Cardiovascular System | 18,458 (15) | 5,041 (27)[a] | 1,295 (27)[abc] | 1,317 (32)[ab] | 2,429 (25)[a] | 13,417 (13) |
| Nervous System | 9,473 (8) | 870 (5)[a] | 219 (5)[a] | 159 (4)[ab] | 492 (5)[a] | 8,603 (8) |
| Respiratory System | 5,948 (5) | 1,745 (9)[a] | 471 (10)[abc] | 588 (14)[ab] | 686 (7) | 4,203 (4) |
| Maternity Care and Delivery | 9,194 (7) | 674 (4)[a] | 212 (4)[ac] | 89 (2)[ab] | 373 (4)[a] | 8,520 (8) |
| Urinary System | 5,094 (4) | 1,441 (8)[a] | 394 (8)[ac] | 175 (4)[b] | 872 (9)[a] | 3,653 (3) |
| **Primary admission diagnostic groups, n (%)** | | | | | | |
| Diseases of the circulatory system | 930,981 (43) | 148,526 (49)[a] | 45,533 (49)[abc] | 21,839 (48)[a] | 81,154 (50)[a] | 782,455 (41) |
| Symptoms; signs; and ill-defined conditions and factors influencing health status | 45,185 (2) | 3,878 (1)[a] | 1,082 (1)[abc] | 440 (1)[ab] | 2,356 (1)[a] | 41,307 (2) |

| Variables | All Cohort (N=2,187,254) | AKI (N= 301,464, %14) | Persistent AKI without renal recovery (N=92,607, %4) | Persistent AKI with renal recovery (N=45,749, %2) | Rapidly Reversed AKI (N=163,108, %8) | No AKI (N= 1,885,790, %86) |
|---|---|---|---|---|---|---|
| Diseases of the digestive system | 176,469 (8) | 18,836 (6)[a] | 5,417 (6)[abc] | 3,033 (7)[ab] | 10,386 (6)[a] | 157,633 (8) |
| Diseases of the genitourinary system | 87,964 (4) | 22,511 (7)[a] | 6,186 (7)[abc] | 3,881 (8)[ab] | 12,444 (8)[a] | 65,453 (3) |
| Diseases of the musculoskeletal system and connective tissue | 24,508 (1) | 1,112 (0)[a] | 393 (0)[ac] | 100 (0)[ab] | 619 (0)[a] | 23,396 (1) |
| Diseases of the skin and subcutaneous tissue | 11,901 (1) | 541 (0)[a] | 178 (0)[ac] | 48 (0)[ab] | 315 (0)[a] | 11,360 (1) |
| Diseases of the respiratory system | 139,520 (6) | 21,093 (7)[a] | 6,936 (7)[ab] | 3,440 (8)[ab] | 10,717 (7)[a] | 118,427 (6) |
| Neoplasms | 30,630 (1) | 3,286 (1)[a] | 1,297 (1)[bc] | 453 (1)[a] | 1,536 (1)[a] | 27,344 (1) |
| Complications of pregnancy; childbirth; and the puerperium | 66,223 (3) | 2,742 (1)[a] | 1,137 (1)[abc] | 270 (1)[ab] | 1,335 (1)[a] | 63,481 (3) |
| Infectious and parasitic diseases | 12,483 (1) | 1,291 (0)[a] | 383 (0)[a] | 190 (0)[a] | 718 (0)[a] | 11,192 (1) |

Abbreviations. SD, standard deviation; IQR, Interquartile Range.

[a] p<0.05 compared to no AKI

[b] p<0.05 compared to Rapidly Reversed AKI

[c] p<0.05 compared to Persistent AKI with renal recovery

[*] Cardiovascular disease was considered if there was a history of congestive heart failure, coronary artery disease of peripheral vascular disease.

[**] Surgical procedures and groupings were identified by considering Current Procedural Terminology (CPT) codes within the range 10004–69990.

**Supplemental Table 2.** Baseline characteristics by trajectory groups in hospitalized adult patients who have been admitted to ICU during hospitalization.

| **Variables** | **All Cohort (N=83,478)** | **AKI (N=24,477, %29)** | **Persistent AKI without renal recovery (N=8,099, %10)** | **Persistent AKI with renal recovery (N=5,222, %6)** | **Rapidly Reversed AKI (N=11,156, %13)** | **No AKI (N=59,001, %71)** |
|---|---|---|---|---|---|---|
| **Preadmission Clinical Characteristics** | | | | | | |
| Age, mean (SD), years | 59 (21) | 64 (21)[a] | 67 (23)[abc] | 63 (19)[a] | 62 (20)[a] | 57 (20) |
| Female sex, n (%) | 36,786 (44) | 11,140 (46)[a] | 3,848 (48)[abc] | 2,316 (44) | 4,976 (45) | 25,646 (43) |
| Race, n (%) | | | | | | |
| African American | 21,143 (25) | 6,941 (28)[a] | 2,209 (27)[a] | 1,500 (29)[a] | 3,232 (29)[a] | 14,202 (24) |
| White | 56,709 (68) | 16,046 (66)[a] | 5,283 (65)[a] | 3,446 (66)[a] | 7,317 (66)[a] | 40,663 (69) |
| Primary Insurance, n (%) | | | | | | |
| Medicare | 36,988 (44) | 13,030 (53)[a] | 4,407 (54)[ab] | 2,831 (54)[ab] | 5,792 (52)[a] | 23,958 (41) |
| Private | 17,920 (21) | 3,996 (16)[a] | 1,229 (15)[ab] | 927 (18)[a] | 1,840 (16)[a] | 13,924 (24) |
| Medicaid | 16,284 (20) | 5,004 (20)[a] | 1,706 (21)[a] | 1,059 (20) | 2,239 (20) | 11,280 (19) |
| Uninsured | 1,990 (2) | 368 (2)[a] | 117 (1)[a] | 68 (1)[a] | 183 (2)[a] | 1622 (3) |
| Other | 7,070 (8) | 1,403 (6)[a] | 483 (6)[ac] | 228 (4)[ab] | 692 (6)[a] | 5,667 (10) |
| Charity | 3,226 (4) | 676 (3)[a] | 157 (2)[ab] | 109 (2)[ab] | 410 (4)[a] | 2,550 (4) |
| **Comorbidities, n (%)** | | | | | | |
| Hypertension, n (%) | 20,536 (25) | 8,527 (35)[a] | 2,562 (32)[abc] | 2,134 (41)[ab] | 3,831 (34)[a] | 12,009 (20) |

| Variables | All Cohort (N=83,478) | AKI (N=24,477, %29) | Persistent AKI without renal recovery (N=8,099, %10) | Persistent AKI with renal recovery (N=5,222, %6) | Rapidly Reversed AKI (N=11,156, %13) | No AKI (N=59,001, %71) |
|---|---|---|---|---|---|---|
| Chronic pulmonary disease, n (%) | 29,538 (35) | 9,908 (40)[a] | 3,164 (39)[ab] | 2,107 (40)[a] | 4,637 (42)[a] | 19,630 (33) |
| Cardiovascular disease, n (%) | 32,221 (39) | 12,245 (50)[a] | 3,939 (49)[ac] | 2,781 (53)[ab] | 5,525 (50)[a] | 19,976 (34) |
| Diabetes mellitus, n (%) | 24,740 (30) | 9,104 (37)[a] | 2,738 (34)[abc] | 1,969 (38)[a] | 4,397 (39)[a] | 15,636 (27) |
| Chronic kidney disease, n (%) | 21,238 (25) | 9,303 (38)[a] | 2,980 (37)[ac] | 2,348 (45)[ab] | 3,975 (36)[a] | 11,935 (20) |
| Moderate/Severe (>= G-Stage 3), n (%) | 10,754 (51) | 4,635 (50)[a] | 1,434 (48)[a] | 1,160 (49) | 2,041 (51) | 6,119 (51) |
| Preadmission estimated glomerular filtration rate (mL/min per 1.73 m2), median (IQR) | 59.61 (41.95, 81.61) | 60.16 (42.32, 83.90)[a] | 61.91 (41.31, 90.34)[ab] | 60.47 (42.94, 80.83) | 59.07 (42.67, 80.35) | 59.21 (41.64, 80.00) |
| Transfer from another hospital, n (%) | 19,723 (24) | 6,627 (27)[a] | 2,516 (31)[abc] | 1,501 (29)[ab] | 2,610 (23) | 13,096 (22) |
| Surgery on admission day, n (%) | 10,399 (12) | 2,636 (11)[a] | 699 (9)[abc] | 680 (13)[ab] | 1,257 (11)[a] | 7,763 (13) |
| Surgery at any time, n (%) | 25,075 (30) | 7,473 (31)[a] | 2,004 (25)[abc] | 2,216 (42)[ab] | 3,253 (29) | 17,602 (30) |
| **Location/Type of surgery**, n (%)** | | | | | | |
| Musculoskeletal System | 3,852 (15) | 730 (10)[a] | 162 (8)[ab] | 204 (9)[a] | 364 (11)[a] | 3,122 (18) |
| Integumentary System | 2,480 (10) | 526 (7)[a] | 147 (7)[a] | 131 (6)[a] | 248 (8)[a] | 1954 (11) |
| Digestive System | 4,340 (17) | 1,750 (23)[a] | 489 (24)[a] | 512 (23)[a] | 749 (23)[a] | 2,590 (15) |
| Cardiovascular System | 6,179 (25) | 2,423 (32)[a] | 662 (33)[a] | 761 (34)[ab] | 1,000 (31)[a] | 3,756 (21) |
| Nervous System | 4,068 (16) | 564 (8)[a] | 129 (6)[ab] | 107 (5)[ab] | 328 (10)[a] | 3,504 (20) |

| **Variables** | **All Cohort (N=83,478)** | **AKI (N=24,477, %29)** | **Persistent AKI without renal recovery (N=8,099, %10)** | **Persistent AKI with renal recovery (N=5,222, %6)** | **Rapidly Reversed AKI (N=11,156, %13)** | **No AKI (N=59,001, %71)** |
|---|---|---|---|---|---|---|
| Respiratory System | 2,752 (11) | 1,011 (14)[a] | 285 (14)[ab] | 346 (16)[ab] | 380 (12)[a] | 1,741 (10) |
| Maternity Care and Delivery | 211 (1) | 59 (1) | 18 (1) | 15 (1) | 26 (1) | 152 (1) |
| Urinary System | 320 (1) | 159 (2)[a] | 42 (2)[ab] | 62 (3)[ab] | 55 (2)[a] | 161 (1) |
| **Primary admission diagnostic groups, n (%)** | | | | | | |
| Diseases of the circulatory system | 43,321 (52) | 14,287 (58)[a] | 4,819 (60)[ab] | 3,106 (59)[a] | 6,362 (57)[a] | 29,034 (49) |
| Symptoms; signs; and ill-defined conditions and factors influencing health status | 2,077 (2) | 546 (2)[a] | 151 (2)[a] | 125 (2) | 270 (2) | 1531 (3) |
| Diseases of the digestive system | 4,067 (5) | 1,380 (6)[a] | 493 (6)[a] | 294 (6)[a] | 593 (5)[a] | 2,687 (5) |
| Diseases of the genitourinary system | 1,115 (1) | 484 (2)[a] | 136 (2)[a] | 122 (2)[a] | 226 (2)[a] | 631 (1) |
| Diseases of the musculoskeletal system and connective tissue | 540 (1) | 34 (0)[a] | 6 (0)[a] | 3 (0)[a] | 25 (0)[a] | 506 (1) |
| Diseases of the skin and subcutaneous tissue | 30 (0) | 5 (0)[a] | 2 (0) | 0 (0) | 3 (0) | 25 (0) |
| Diseases of the respiratory system | 12,263 (15) | 3,226 (13)[a] | 1,110 (14)[ac] | 6,07 (12)[ab] | 1,509 (14)[a] | 9,037 (15) |
| Neoplasms | 435 (1) | 86 (0)[a] | 40 (0) | 13 (0)[a] | 33 (0)[a] | 349 (1) |
| Complications of pregnancy; childbirth; and the puerperium | 405 (0) | 80 (0)[a] | 27 (0) | 15 (0) | 38 (0)[a] | 325 (1) |
| Infectious and parasitic diseases | 94 (0) | 18 (0)[a] | 7 (0) | 1 (0) | 10 (0) | 76 (0) |

Abbreviations. SD, standard deviation; IQR, Interquartile Range; KRT, kidney replacement therapy.
[a] $p<0.05$ compared to no AKI
[b] $p<0.05$ compared to Rapidly Reversed AKI
[c] $p<0.05$ compared to Persistent AKI with renal recovery
[*] Cardiovascular disease was considered if there was a history of congestive heart failure, coronary artery disease of peripheral vascular disease.
[**] Surgical procedures and groupings were identified by considering Current Procedural Terminology (CPT) codes within the range 10004–69990.

**Supplemental Table 3.** Baseline characteristics by trajectory groups in hospitalized adult patients who have not been admitted to ICU at any time during hospitalization.

| Variables | All Cohort (N=2,103,776) | AKI (N=276,987, %13) | Persistent AKI without renal recovery (N=84,508, %4) | Persistent AKI with renal recovery (N=40,527, %2) | Rapidly Reversed AKI (N=151,952, %7) | No AKI (N=1,826,789, %87) |
|---|---|---|---|---|---|---|
| **Preadmission Clinical Characteristics** | | | | | | |
| Age, mean (SD), years | 59 (22) | 69 (24)[a] | 74 (27)[abc] | 69 (22)[ab] | 67 (23)[a] | 58 (22) |
| Female sex, n (%) | 1,130,319 (54) | 145,981 (53)[a] | 48,047 (57)[abc] | 20,719 (51)[a] | 77,215 (51)[a] | 984,338 (54) |
| Race, n (%) | | | | | | |
| African American | 411,448 (20) | 58,284 (21)[a] | 16,839 (20)[abc] | 8,633 (21)[a] | 32,812 (22)[a] | 35,3164 (19) |
| White | 1,419,457 (67) | 183,749 (66)[a] | 56,636 (67)[abc] | 26,554 (66)[a] | 100,559 (66)[a] | 1,235,708 (68) |
| Primary Insurance, n (%) | | | | | | |
| Medicare | 363,396 (17) | 65,643 (24)[a] | 20,527 (24)[abc] | 9,168 (23)[ab] | 35,948 (24)[a] | 297,753 (16) |
| Private | 691,646 (33) | 78,012 (28 [a]) | 22,416 (27)[ab] | 10,915 (27)[ab] | 44,681 (29)[a] | 613,634 (34) |
| Medicaid | 184,701 (9) | 24,728 (9)[a] | 6,771 (8)[abc] | 3,756 (9)[a] | 14,201 (9)[a] | 159,973 (9) |
| Uninsured | 68,893 (3) | 5,711 (2)[a] | 1,441 (2)[ab] | 631 (2)[ab] | 3,639 (2)[a] | 63,182 (3) |
| Other | 725,129 (34) | 97,299 (35)[a] | 32,224 (38)[ab] | 15,458 (38)[ab] | 49,617 (33)[a] | 627,830 (34) |
| Charity | 70,008 (3) | 5,594 (2)[a] | 1,129 (1)[ab] | 599 (1)[ab] | 3,866 (3)[a] | 64,414 (4) |
| **Comorbidities, n (%)** | | | | | | |
| Hypertension, n (%) | 342,009 (16) | 81,325 (29)[a] | 22,278 (26)[abc] | 13,919 (34)[ab] | 45,128 (30)[a] | 260,684 (14) |

| Variables | All Cohort (N=2,103,776) | AKI (N=276,987, %13) | Persistent AKI without renal recovery (N=84,508, %4) | Persistent AKI with renal recovery (N=40,527, %2) | Rapidly Reversed AKI (N=151,952, %7) | No AKI (N=1,826,789, %87) |
|---|---|---|---|---|---|---|
| Chronic pulmonary disease, n (%) | 574,934 (27) | 87,059 (31)[a] | 23,576 (28)[abc] | 13,653 (34)[ab] | 49,830 (33)[a] | 487,875 (27) |
| Cardiovascular disease, n (%) | 530,682 (25) | 103,638 (37)[a] | 28,827 (34)[abc] | 17,209 (42)[ab] | 57,602 (38)[a] | 427,044 (23) |
| Diabetes mellitus, n (%) | 575,048 (27) | 104,080 (38)[a] | 27,992 (33)[abc] | 16,165 (40)[a] | 59,923 (39)[a] | 470,968 (26) |
| Chronic kidney disease, n (%) | 454,402 (22) | 102,198 (37)[a] | 25,002 (30)[abc] | 18,892 (47)[ab] | 58,304 (38)[a] | 352,204 (19) |
| Moderate/Severe (>= G-Stage 3), n (%) | 252,890 (56) | 60,194 (59)[a] | 13,588 (54)[bc] | 11,556 (61)[a] | 35,050 (60)[a] | 192,696 (55) |
| Preadmission estimated glomerular filtration rate (mL/min per 1.73 m2), median (IQR) | 56.47 (40.82, 74.48) | 53.74 (37.39, 73.60)[a] | 56.52 (38.11, 80.51)[bc] | 52.11 (36.09, 71.77)[ab] | 53.39 (37.54, 71.78)[a] | 57.18 (41.77, 74.71) |
| Transfer from another hospital, n (%) | 152,884 (7) | 28,078 (10)[a] | 102,76 (12)[ab] | 5,117 (13)[ab] | 12,685 (8)[a] | 124,806 (7) |
| Surgery on admission day, n (%) | 33,022 (2) | 3,505 (1)[a] | 833 (1)[abc] | 391 (1)[ab] | 2,281 (2)[a] | 29,517 (2) |
| Surgery at any time, n (%) | 99,738 (5) | 11,387 (4)[a] | 2,807 (3)[abc] | 1,939 (5)[b] | 6,641 (4)[a] | 88,351 (5) |
| **Location/Type of surgery**, n (%)** | | | | | | |
| Musculoskeletal System | 26,265 (26) | 2,245 (20)[a] | 570 (20)[a] | 347 (18)[a] | 1,328 (20)[a] | 24,020 (27) |
| Integumentary System | 9,773 (10) | 972 (9)[a] | 283 (10)[bc] | 137 (7)[a] | 552 (8)[a] | 8,801 (10) |
| Digestive System | 21,345 (21) | 1,957 (17)[a] | 375 (13)[abc] | 322 (17)[a] | 1,260 (19)[a] | 19,388 (22) |
| Cardiovascular System | 12,279 (12) | 2,618 (23)[a] | 633 (23)[ac] | 556 (29)[ab] | 1,429 (22)[a] | 9,661 (11) |
| Nervous System | 5,405 (5) | 306 (3)[a] | 90 (3)[a] | 52 (3)[a] | 164 (2)[a] | 5,099 (6) |

| Variables | All Cohort (N=2,103,776) | AKI (N=276,987, %13) | Persistent AKI without renal recovery (N=84,508, %4) | Persistent AKI with renal recovery (N=40,527, %2) | Rapidly Reversed AKI (N=151,952, %7) | No AKI (N=1,826,789, %87) |
|---|---|---|---|---|---|---|
| Respiratory System | 3,196 (3) | 734 (6)[a] | 186 (7)[abc] | 242 (12)[ab] | 306 (5)[a] | 2,462 (3) |
| Maternity Care and Delivery | 8,983 (9) | 615 (5)[a] | 194 (7)[ac] | 74 (4)[a] | 347 (5)[a] | 8,368 (9) |
| Urinary System | 4,774 (5) | 1,282 (11)[a] | 352 (13)[ac] | 113 (6)[ab] | 817 (12)[a] | 3,492 (4) |
| **Primary admission diagnostic groups, n (%)** | | | | | | |
| Diseases of the circulatory system | 887,660 (42) | 134,239 (48)[a] | 40,714 (48)[abc] | 18,733 (46)[ab] | 74,792 (49)[a] | 75,3421 (41) |
| Symptoms; signs; and ill-defined conditions and factors influencing health status | 43,108 (2) | 3,332 (1)[a] | 931 (1)[abc] | 315 (1)[ab] | 2,086 (1)[a] | 39,776 (2) |
| Diseases of the digestive system | 172,402 (8) | 17,456 (6)[a] | 4,924 (6)[abc] | 2,739 (7)[a] | 9,793 (6)[a] | 154,946 (8) |
| Diseases of the genitourinary system | 86,849 (4) | 22,027 (8)[a] | 6,050 (7)[abc] | 3,759 (9)[ab] | 12,218 (8)[a] | 64,822 (4) |
| Diseases of the musculoskeletal system and connective tissue | 23,968 (1) | 1,078 (0)[a] | 387 (0)[ac] | 97 (0)[ab] | 594 (0)[a] | 22,890 (1) |
| Diseases of the skin and subcutaneous tissue | 11,871 (1) | 536 (0)[a] | 176 (0)[ac] | 48 (0)[ab] | 312 (0)[a] | 11,335 (1) |
| Diseases of the respiratory system | 127,257 (6) | 17,867 (6)[a] | 5,826 (7)[ab] | 2,833 (7)[ab] | 9,208 (6) | 109,390 (6) |
| Neoplasms | 30,195 (1) | 3,200 (1)[a] | 1,257 (1)[bc] | 440 (1)[a] | 1,503 (1)[a] | 26,995 (1) |

| **Variables** | **All Cohort (N=2,103,776)** | **AKI (N=276,987, %13)** | **Persistent AKI without renal recovery (N=84,508, %4)** | **Persistent AKI with renal recovery (N=40,527, %2)** | **Rapidly Reversed AKI (N=151,952, %7)** | **No AKI (N=1,826,789, %87)** |
|---|---|---|---|---|---|---|
| Complications of pregnancy; childbirth; and the puerperium | 65,818 (3) | 2,662 (1)[a] | 1,110 (1)[abc] | 255 (1)[ab] | 1,297 (1)[a] | 63,156 (3) |
| Infectious and parasitic diseases | 12,389 (1) | 1,273 (0)[a] | 376 (0)[a] | 189 (0)[a] | 708 (0)[a] | 11,116 (1) |

Abbreviations. SD, standard deviation; IQR, Interquartile Range; KRT, kidney replacement therapy.

[a] $p<0.05$ compared to no AKI

[b] $p<0.05$ compared to Rapidly Reversed AKI

[c] $p<0.05$ compared to Persistent AKI with renal recovery

* Cardiovascular disease was considered if there was a history of congestive heart failure, coronary artery disease of peripheral vascular disease.

** Surgical procedures and groupings were identified by considering Current Procedural Terminology (CPT) codes within the range 10004–69990.

**Supplemental Table 4.** CKD and AKI characteristics of the cohort based on phenotyping algorithm.

| | **n (%)** |
|---|---|
| **Overall number of encounters** | 2,187,254 |
| **Preadmission CKD, n (%)** | |
| Insufficient Data (No CKD with warning) | 2,309 (0.1) |
| No CKD | 1,709,305 (78) |
| CKD | 475,640 (22) |
| CKD by Medical History | 376,838 (79) |
| CKD by Creatinine Criteria | 86,665 (18) |
| CKD after kidney transplant | 12,137 (2) |
| **Reference creatinine determination method, n (%)** | |
| **Among encounters with CKD** | |
| Admission creatinine | 222,053 (46) |
| Minimum creatinine in the 7 days prior to admission | 45,279 (10) |
| Median creatinine in 8-365 days prior to admission | 198,796 (42) |
| First creatinine | 9,512 (2) |
| **Among encounters with no CKD[a]** | |
| Admission creatinine | 829,130 (48) |
| Minimum creatinine in the 7 days prior to admission | 132,584 (8) |
| Median creatinine in 8-365 days prior to admission | 428,070 (25) |
| Estimated creatinine (CKD-EPI)[b] | 240,129 (14) |
| First creatinine | 81,701 (5) |
| **Eligible number of encounters, n** | |
| **No AKI during hospitalization, n (%)** | 1,885,790 (86) |
| **AKI during hospitalization, n (%)** | 301,464 (14) |
| Reference serum creatinine, median ($25^{th}$, $75^{th}$) | 0.82 (0.61, 1.01) |
| Reference serum creatinine, mean (SD) | 0.91 (0.5) |
| **Maximum AKI Stage, n (%)** | |
| Stage 1 | 190,244 (63) |
| Stage 2 | 63,519 (21) |
| Stage 3 (with or without KRT) | 47,701 (16) |
| KRT, n (%) | 1,994 |
| Number of days on KRT, median ($25^{th}$, $75^{th}$) | 8 (4,12) |
| Recurrent AKI, n (%) | 23,938 (0.7) |
| AKI duration, days, median ($25^{th}$, $75^{th}$) | 2 (1,4) |
| **AKI trajectories, n (%)** | |
| Rapidly reversed AKI | 163,108 (69) |
| Persistent AKI | 138,356 (31) |

Abbreviations: AKI, acute kidney injury; CKD, chronic kidney disease; eGFR, estimated glomerular filtration rate; KRT, kidney replacement therapy.
[a] Includes group with No CKD with warning due to insufficient data.
[b] Algorithm calculated estimated creatinine by back calculation from the 2021 CKD-EPI refit without race.

**Supplemental Table 5.** Detailed CKD characteristics of the cohort based on phenotyping algorithm.

| | n (%) |
|---|---|
| **Overall Patient Encounters** | 2,187,254 |
| Insufficient Data (No CKD with warning) | 2,309 (0.1) |
| **No CKD** | 1,709,305 (78) |
| No CKD by Medical History or Creatinine Criteria and no recent AKI episode[a] | 1,666,963 (98) |
| No CKD by Medical History or Creatinine Criteria, Recovered recent AKI on Admission | 31,072 (2) |
| No CKD by Medical History or Creatinine Criteria, Non-recovered recent AKI (AKD) on admission | 11,270 (0.6) |
| **CKD** | |
| **CKD by Medical History** | 376,838 (80) |
| CKD by Medical History and no recent AKI episode[a] | 283,148 (75) |
| CKD by Medical History, Recovered recent AKI on Admission | 35,814 (10) |
| CKD by Medical History, Non-recovered recent AKI (AKD)on Admission | 57,876 (15) |
| **CKD by Creatinine Criteria** | 86,665 (18) |
| CKD by Creatinine Criteria and no recent AKI episode[a] | 76,900 (89) |
| CKD by Creatinine Criteria, Recovered recent AKI on Admission | 5,846 (7) |
| CKD by Creatinine Criteria, Non-recovered recent AKI (AKD)on Admission | 3,919 (4) |
| **CKD after kidney transplant** | 12,137 (2) |
| CKD after kidney transplant and no recent AKI episode[a] | 8,044 (66) |
| CKD after kidney transplant, Recovered recent AKI on Admission | 1,322 (11) |
| CKD after kidney transplant, Non-recovered recent AKI (AKD)on Admission | 2,771 (23) |
| **CKD G Stages among all encounters with CKD** | |
| eGFR, ml/min/1.73m$^2$, median (25$^{th}$, 75$^{th}$) | 56.59 (40.86, 74.75) |
| eGFR, ml/min/1.73m$^2$), mean (SD) | 59.37 (25.56) |
| G1 (eGFR ≥ 90 ml/min/1.73m$^2$) | 65,451 (14) |
| G2 (90>eGFR ≥ 60 ml/min/1.73m$^2$) | 146,545 (31) |
| G3a (60>eGFR ≥ 45 ml/min/1.73m$^2$) | 115,400 (24) |
| G3b (45>eGFR ≥ 30 ml/min/1.73m$^2$) | 92,686 (19) |
| G4 (30>eGFR ≥ 15 ml/min/1.73m$^2$) | 46,816 (10) |
| G5 (eGFR < 15 ml/min/1.73m$^2$) | 8,742 (2) |
| No staging can be done | NA |

[a] Recent AKI episode defined by the presence of ICD9 or 10 codes documented in EHR in the three months prior to admission.

**Supplemental Table 6.** Summary of number of serum creatinine measurements in first two days of AKI onset.

| Number of serum creatinine measurements | AKI | Persistent AKI with no renal recovery | Persistent AKI with renal recovery | Rapidly Reversed AKI |
|---|---|---|---|---|
| Day 1 of AKI onset | | | | |
| Median (25th, 75th) | 1 (1, 2) | 1 (1, 2) | 1 (1, 2) | 1 (1, 2) |
| Mean (SD) | 1.57 (1.24) | 1.60 (1.34) | 1.78 (1.73) | 1.50 (0.98) |
| Day 2 of AKI onset | | | | |
| Median (25th, 75th) | 1 (1, 1) | 1 (0, 1) | 1 (1, 1) | 1 (1, 1) |
| Mean (SD) | 1.03 (0.99) | 0.97 (1.04) | 1.27 (1.27) | 1.00 (0.86) |
| First two days after AKI onset | | | | |
| Median (25th, 75th) | 2 (2, 3) | 2 (2, 3) | 2 (2, 3) | 2 (2, 3) |
| Mean (SD) | 2.60 (1.76) | 2.56 (1.84) | 3.05 (2.31) | 2.49 (1.49) |

Abbreviations. AKI, acute kidney injury; SD, standard deviation

**Supplemental Table 7.** Renal outcomes after hospital discharge by trajectories of AKI in all cohort who discharged alive.

| Variables | All Cohort (N=2,167,713) | AKI (N=289,798) | Persistent AKI without renal recovery (N=83,403) | Persistent AKI with renal recovery (N=44,754) | Rapidly Reversed AKI (N=161,641) | No AKI (N=1,877,915) |
|---|---|---|---|---|---|---|
| **Ninety-day outcomes** | | | | | | |
| Dead, n (%) | 20,497 (1) | 5,800 (2)[a] | 1,766 (2)[abc] | 1,268 (3)[ab] | 2,766 (2)[a] | 14,697 (1) |
| Readmission, n (%) | 748,297 (35) | 112,868 (39)[a] | 29,024 (35)[abc] | 19,897 (44)[ab] | 63,947 (40)[a] | 635,429 (34) |
| No KRT, n (%) | 2,166,485 (100) | 288,570 (100)[a] | 82,580 (99)[ab] | 44,377 (99)[ab] | 161,613 (100)[a] | 1,877,915 (100) |
| New KRT, n (%) | 752 (0) | 289 (0)[a] | 80 (0)[ac] | 81 (0)[ab] | 128 (0)[a] | 463 (0) |
| No CKD, n (%) | 1,698,477 (78) | 182,581 (63)[a] | 58,703 (70)[abc] | 23,985 (54)[ab] | 99,893 (62)[a] | 1,515,896 (81) |
| New CKD/ESRD, n (%) | 29,002 (1) | 10,766 (4)[a] | 3,836 (5)[abc] | 1,868 (4)[ab] | 5,062 (3)[a] | 18,236 (1) |
| **One-year outcomes** | | | | | | |
| Dead, n (%) | 47,156 (2) | 12,252 (4)[a] | 3,276 (4)[ac] | 2,677 (6)[ab] | 6,299 (4)[a] | 34,904 (2) |
| Readmission, n (%) | 1,018,184 (47) | 148,727 (49)[a] | 37,867 (41)[abc] | 24,969 (55)[ab] | 85,891 (53)[a] | 869,457 (46) |
| No KRT on index admission, n (%) | 2,166,485 (100) | 28,8570 (100)[a] | 82,580 (99)[ab] | 44,377 (99)[ab] | 161,613 (100)[a] | 1,877,915 (100) |
| New KRT, n (%) | 1,788 (0) | 608 (0)[a] | 158 (0)[ac] | 154 (0)[ab] | 296 (0)[a] | 1,180 (0) |
| No CKD on index admission, n (%) | 1,698,477 (78) | 182,581 (63)[a] | 58,703 (70)[ab] | 23,985 (54)[ab] | 99,893 (62)[a] | 1,515,896 (81) |
| New CKD/ESRD, n (%) | 77,031 (4) | 23,959 (8)[a] | 8,110 (10)[abc] | 3,713 (8)[ab] | 12,136 (8)[a] | 53,072 (3) |
| CKD on index admission, n (%) | 469,236 (22) | 107,217 (37)[a] | 24,700 (30)[abc] | 20,769 (46)[ab] | 61,748 (38)[a] | 362,019 (19) |
| CKD progression, n (%) | 26,456 (6) | 13,253 (12)[a] | 2,604 (11)[abc] | 2,786 (13)[ab] | 7,863 (13)[a] | 13,293 (4) |

Abbreviations. CKD, chronic kidney disease; KRT, kidney replacement therapy; ESRD, end stage of renal disease.

[a] $p<0.05$ compared to no AKI

[b] $p<0.05$ compared to Rapidly Reversed AKI

[c] $p<0.05$ compared to Persistent AKI with renal recovery

**Supplemental Table 8.** Renal outcomes after hospital discharge by trajectories of AKI in hospitalized adult patients who have been admitted to ICU during hospitalization and discharged alive.

| **Variables** | **All Cohort (N=74,015)** | **AKI (N=18,651)** | **Persistent AKI without renal recovery (N=3,562)** | **Persistent AKI with renal recovery (N=4,719)** | **Rapidly Reversed AKI (N=10,370)** | **No AKI (N=55,364)** |
|---|---|---|---|---|---|---|
| **Ninety-day outcomes** | | | | | | |
| Dead, n (%) | 4,033 (5) | 1,616 (9)[a] | 453 (13)[abc] | 421 (9)[ab] | 742 (7)[a] | 2,417 (4) |
| Readmission, n (%) | 6,510 (9) | 2,199 (12)[a] | 376 (11)[abc] | 595 (13)[ab] | 1,228 (12)[a] | 43,11 (8) |
| No KRT, n (%) | 73,578 (99) | 18,214 (98)[a] | 3,322 (93)[abc] | 4,524 (96)[ab] | 10,368 (100) | 55,364 (100) |
| New KRT, n (%) | 75 (0) | 39 (0)[a] | 11 (0)[a] | 13 (0)[a] | 15 (0) | 36 (0) |
| No CKD, n (%) | 55,870 (75) | 11,580 (62)[a] | 2,295 (64)[ac] | 2,612 (55)[ab] | 6,673 (64)[a] | 44,290 (80) |
| New CKD/ESRD, n (%) | 1260 (2) | 687 (4)[a] | 189 (5)[abc] | 186 (4)[a] | 312 (3)[a] | 573 (1) |
| **One-year outcomes** | | | | | | |
| Dead, n (%) | 8,140 (11) | 3,145 (17)[a] | 761 (21)[abc] | 874 (19)[ab] | 1,510 (15)[a] | 4,995 (9) |
| Readmission, n (%) | 11,038 (13) | 3,549 (14)[a] | 575 (7)[bc] | 918 (18)[a] | 2,056 (18)[a] | 7,489 (13) |
| No KRT on index admission, n (%) | 73,578 (99) | 18,214 (98)[a] | 3,322 (93)[abc] | 4,524 (96)[ab] | 10,368 (100) | 55,364 (100) |
| New KRT, n (%) | 149 (0) | 65 (0)[a] | 17 (0)[a] | 23 (0)[a] | 25 (0) | 84 (0) |
| No CKD on index admission, n (%) | 55,870 (75) | 11,580 (62)[a] | 2,295 (64)[ac] | 2,612 (55)[ab] | 6,673 (64)[a] | 44,290 (80) |
| New CKD/ESRD, n (%) | 2,865 (4) | 1,350 (7)[a] | 318 (9)[ab] | 366 (8)[ab] | 666 (6)[a] | 1,515 (3) |
| CKD on index admission, n (%) | 18,145 (25) | 7,071 (38)[a] | 1,267 (36)[ac] | 2,107 (45)[ab] | 3,697 (36)[a] | 11,074 (20) |
| CKD progression, n (%) | 1,049 (6) | 711 (10)[a] | 111 (9) [abc] | 233 (11) [ab] | 367 (10)[a] | 338 (3) |

Abbreviations. CKD, chronic kidney disease; KRT, kidney replacement therapy; ESRD, end stage of renal disease.

[a] $p<0.05$ compared to no AKI

[b] $p<0.05$ compared to Rapidly Reversed AKI

[c] $p<0.05$ compared to Persistent AKI with renal recovery

**Supplemental Table 9.** Renal outcomes after hospital discharge by trajectories of AKI in hospitalized adult patients who have not been admitted to ICU at any time during hospitalization and discharged alive.

| Variables | All Cohort (N=2,093,698) | AKI (N=271,147) | Persistent AKI without renal recovery (N=79,841) | Persistent AKI with renal recovery (N=40,035) | Rapidly Reversed AKI (N=151,271) | No AKI (N=1,822,551) |
|---|---|---|---|---|---|---|
| **Ninety-day outcomes** | | | | | | |
| Dead, n (%) | 16,464 (1) | 4,184 (2)[a] | 1,313 (2)[abc] | 847 (2)[ab] | 2,024 (1)[a] | 12,280 (1) |
| Readmission, n (%) | 718,364 (34) | 104,775 (39)[a] | 27,518 (34)[abc] | 17,831 (45)[ab] | 59,426 (39)[a] | 613,589 (34) |
| No KRT, n (%) | 2,092,907 (100) | 270,356 (100)[a] | 79,258 (99)[abc] | 39,853 (100)[ab] | 151,245 (100)[a] | 1,822,551 (100) |
| New KRT, n (%) | 677 (0) | 250 (0)[a] | 69 (0)[ac] | 68 (0)[ab] | 113 (0)[a] | 427 (0) |
| No CKD, n (%) | 1,642,607 (78) | 171,001 (63)[a] | 56,408 (71)[abc] | 21,373 (53)[ab] | 93,220 (62)[a] | 1,471,606 (81) |
| New CKD/ESRD, n (%) | 27,742 (1) | 10,079 (4)[a] | 3,647 (5)[abc] | 1,682 (4)[ab] | 4,750 (3)[a] | 17,663 (1) |
| **One-year outcomes** | | | | | | |
| Dead, n (%) | 39,016 (2) | 9,107 (3)[a] | 2,515 (3)[ac] | 1,803 (5)[ab] | 4,789 (3)[a] | 29,909 (2) |
| Readmission, n (%) | 977,619 (46) | 138,141 (50)[a] | 35,943 (43)[abc] | 22,323 (55)[ab] | 79,875 (53)[a] | 839,478 (46) |
| No KRT on index admission, n (%) | 2,092,907 (100) | 270,356 (100)[a] | 79,258 (99)[abc] | 39,853 (100)[ab] | 151,245 (100[a]) | 1,822,551 (100) |
| New KRT, n (%) | 1639 (0) | 543 (0)[a] | 141 (0)[ac] | 131 (0)[ab] | 271 (0)[a] | 1,096 (0) |
| No CKD on index admission, n (%) | 1,642,607 (78) | 171,001 (63)[a] | 56,408 (71)[abc] | 21,373 (53)[ab] | 93,220 (62)[a] | 1,471,606 (81) |
| New CKD/ESRD, n (%) | 74,166 (4) | 22,609 (8)[a] | 7,792 (10)[abc] | 3,347 (8)[ab] | 11,470 (8)[a] | 51,557 (3) |
| CKD on index admission, n (%) | 451,091 (22) | 100,146 (37)[a] | 23,433 (29)[abc] | 18,662 (47)[ab] | 58,051 (38)[a] | 350,945 (19) |
| CKD progression, n (%) | 25,497 (6) | 12,542 (13)[a] | 2,493 (11)[abc] | 2,553 (14)[ab] | 7496 (13)[a] | 12,955 (4) |

Abbreviations. CKD, chronic kidney disease; KRT, kidney replacement therapy; ESRD, end stage of renal disease.

[a] $p<0.05$ compared to no AKI

[b] $p<0.05$ compared to Rapidly Reversed AKI

[c] p<0.05 compared to Persistent AKI with renal recovery

**Supplemental Table 10.** Renal characteristics, resource utilization, and hospital outcomes during entire hospitalization by trajectories of AKI in hospitalized adult patients who have been admitted to ICU during hospitalization.

| Variables | All Cohort (N=83,478) | AKI (N=24,477) | Persistent AKI without renal recovery (N=8,099) | Persistent AKI with renal recovery (N=5,222) | Rapidly Reversed AKI (N=11,156) | No AKI (N=59,00) |
|---|---|---|---|---|---|---|
| **Renal characteristics during entire hospitalization** | | | | | | |
| Worst AKI Staging, n(%) | | | | | | |
| Stage 1 | 12,917 (15) | 12,917 (53)[a] | 2,213 (27)[abc] | 1,903 (36)[ab] | 8,801 (79)[a] | 0 (0) |
| Stage 2 | 5,760 (7) | 5,760 (24)[a] | 2,356 (29)[ab] | 1,756 (34)[ab] | 1,648 (15)[a] | 0 (0) |
| Stage 3 | 4,782 (6) | 4,782 (20)[a] | 2,739 (34)[ab] | 1,338 (26)[ab] | 705 (6)[a] | 0 (0) |
| Stage 3 with KRT | 1,018 (1) | 1,018 (4) | 791 (10)[abc] | 225 (4)[ab] | 2 (0) | 0 (0) |
| AKI duration, days, median (IQR) | 0 (0, 1) | 2 (1, 5) | 5 (3, 10)[bc] | 5 (3, 8)[b] | 1 (1, 2) | 0 (0, 0) |
| Recurrent AKI, n (%) | 3,830 (5) | 3,830 (16)[a] | 1,580 (20)[abc] | 1,350 (26)[ab] | 900 (8)[a] | 0 (0) |
| No renal recovery at discharge/death, n (%) | 13,887 (17) | 10,250 (42)[a] | 8,099 (100)[abc] | 503 (10)[ab] | 1,648 (15)[a] | 3,637 (6) |
| **Resource utilization during entire hospitalization** | | | | | | |
| Hospital days, median (IQR) | 6 (3, 12) | 9 (4, 17)[a] | 7 (3, 15)[abc] | 16 (9, 27)[ab] | 8 (4, 15)[a] | 5 (2, 9) |
| Mechanical Ventilation, n (%) | 26,238 (31) | 11,041 (45)[a] | 4,554 (56)[abc] | 2,701 (52)[ab] | 3,786 (34)[a] | 15,197 (26) |
| Vasopressor or inotropes used, n (%) | 34,544 (41) | 13,685 (56)[a] | 5,260 (65)[abc] | 3,398 (65)[ab] | 5,027 (45)[a] | 20,859 (35) |
| **Hospital disposition, n(%)** | | | | | | |
| Hospital mortality | 9,463 (11) | 5,826 (24)[a] | 4,537 (56)[abc] | 503 (10) | 786 (7) | 3,637 (6) |
| Another hospital, LTAC, SNF, Hospice | 17,635 (21) | 6,529 (27)[a] | 1,614 (20)[abc] | 2,031 (39) | 2,884 (26) | 11,106 (19) |

| Variables | All Cohort (N=83,478) | AKI (N=24,477) | Persistent AKI without renal recovery (N=8,099) | Persistent AKI with renal recovery (N=5,222) | Rapidly Reversed AKI (N=11,156) | No AKI (N=59,00) |
|---|---|---|---|---|---|---|
| Home/Rehab | 56,380 (68) | 12,122 (50)[a] | 1,948 (24)[bc] | 2,688 (51) | 7,486 (67) | 44,258 (75) |
| **30-day outcomes (among survivors), n (%)** | | | | | | |
| Death in 30 days of discharge | 1,384 (2) | 541 (3)[a] | 193 (5)[abc] | 110 (2) | 238 (2) | 843 (2) |
| Trajectory group for encounter with readmission within 30 days of discharge | | | | | | |
| Persistent AKI without renal recovery | 505 (14) | 240 (19)[a] | 73 (32) | 68 (19) | 99 (14) | 265 (11) |
| Persistent AKI with renal recovery | 323 (9) | 164 (13)[a] | 34 (15) | 55 (16) | 75 (11) | 159 (6) |
| Rapidly Reversed AKI | 727 (19) | 304 (24)[a] | 47 (21) | 74 (21) | 183 (26) | 423 (17) |
| No AKI | 2,178 (58) | 564 (44)[a] | 74 (32) | 154 (44) | 336 (48) | 1,614 (66) |
| **Other complications during entire hospitalization** | | | | | | |
| Venous Thromboembolism, n (%) | 9,967 (12) | 4,134 (17)[a] | 1,387 (17)[abc] | 1,180 (23)[ab] | 1,567 (14) | 5,833 (10) |
| Sepsis, n (%) | 23,331 (28) | 11,778 (48)[a] | 4,539 (56)[ab] | 3,000 (57)[ab] | 4,239 (38)[a] | 11,553 (20) |
| Cardiovascular complication, n (%) | 26,507 (32) | 12,860 (53)[a] | 5,009 (62)[ab] | 3,138 (60)[ab] | 4,713 (42)[a] | 13,647 (23) |
| Thirty-day mortality, n (%) | 10,347 (12) | 5,983 (24)[a] | 4,472 (55)[abc] | 541 (10)[ab] | 970 (9)[a] | 4,364 (7) |
| One-year mortality, n (%) | 17,598 (21) | 8,967 (37)[a] | 5,294 (65)[abc] | 1,377 (26)[ab] | 2,296 (21)[a] | 8,631 (15) |

| **Variables** | **All Cohort (N=83,478)** | **AKI (N=24,477)** | **Persistent AKI without renal recovery (N=8,099)** | **Persistent AKI with renal recovery (N=5,222)** | **Rapidly Reversed AKI (N=11,156)** | **No AKI (N=59,00)** |
|---|---|---|---|---|---|---|
| Three-year mortality, n (%) | 21,170 (25) | 10,192 (42)[a] | 5,506 (68)[abc] | 1,704 (33)[ab] | 2,982 (27)[a] | 10,978 (19) |

Abbreviations. SD, standard deviation; IQR, interquartile range; KRT, kidney replacement therapy; ICU, intensive care unit; LTAC, long-term acute care hospital; SNF, skilled nursing facility; NA, not applicable.

[a] p<0.05 compared to no AKI

[b] p<0.05 compared to Rapidly Reversed AKI

[c] p<0.05 compared to Persistent AKI with renal recovery

**Supplemental Table 11.** Renal characteristics, resource utilization, and hospital outcomes during entire hospitalization by trajectories of AKI in hospitalized adult patients who have not been admitted to ICU any time during hospitalization.

| Variables | All Cohort (N=2,103,776) | AKI (N=276,987) | Persistent AKI without renal recovery (N=84,508) | Persistent AKI with renal recovery (N=40,527) | Rapidly Reversed AKI (N=151,952) | No AKI (N=1,826,789) |
|---|---|---|---|---|---|---|
| **Renal characteristics during entire hospitalization** | | | | | | |
| Worst AKI Staging, n (%) | | | | | | |
| Stage 1 | 177,327 (8) | 177,327 (64)[a] | 35,969 (43)[abc] | 19,076 (47)[ab] | 122,282 (80)[a] | 0 (0) |
| Stage 2 | 57,759 (3) | 57,759 (21)[a] | 25,588 (30)[abc] | 11,470 (28)[ab] | 20,701 (14)[a] | 0 (0) |
| Stage 3 | 40,925 (2) | 40,925 (15)[a] | 22,196 (26)[ab] | 9,786 (24)[ab] | 8,943 (6)[a] | 0 (0) |
| Stage 3 with KRT | 976 (0) | 976 (0)[a] | 755 (1)[ab] | 195 (0)[ab] | 26 (0)[a] | 0 (0) |
| AKI duration, days, median (IQR) | 0 (0, 0) | 2 (1, 4) | 5 (3, 7)[b] | 4 (3, 6)[b] | 1 (1, 2) | 0 (0, 0) |
| Recurrent AKI, n (%) | 20,108 (1) | 20,108 (7)[a] | 7,879 (9)[abc] | 6,179 (15)[ab] | 6,050 (4)[a] | 0 (0) |
| No renal recovery at discharge/death, n (%) | 12,1753 (6) | 117,515 (42)[a] | 84,508 (100)[abc] | 492 (1)[a] | 32,515 (21)[a] | 4238 (0) |
| **Resource utilization during entire hospitalization** | | | | | | |
| Hospital days, median (IQR) | 3 (1, 5) | 5 (3, 9)[a] | 5 (3, 9)[abc] | 10 (6, 17)[ab] | 4 (2, 8)[a] | 2 (1, 5) |
| Mechanical Ventilation, n (%) | 58,228 (3) | 26,979 (10)[a] | 11,177 (13)[abc] | 6,607 (16)[ab] | 9,195 (6)[a] | 31,249 (2) |
| Vasopressor or inotropes used, n (%) | 91,655 (4) | 18,209 (7)[a] | 6,400 (8)[abc] | 3,478 (9)[ab] | 8,331 (5)[a] | 73,446 (4) |
| Hospital disposition, n(%) | | | | | | |
| Hospital mortality | 1,0078 (0) | 5,840 (2)[a] | 4,667 (6)[abc] | 492 (1)[ab] | 681 (0)[a] | 4,238 (0) |

| Variables | All Cohort (N=2,103,776) | AKI (N=276,987) | Persistent AKI without renal recovery (N=84,508) | Persistent AKI with renal recovery (N=40,527) | Rapidly Reversed AKI (N=151,952) | No AKI (N=1,826,789) |
|---|---|---|---|---|---|---|
| Another hospital, LTAC, SNF, Hospice | 20,3926 (10) | 56,997 (21)[a] | 25,006 (30)[abc] | 10,992 (27)[ab] | 20,999 (14)[a] | 146,929 (8) |
| Home/Rehab | 1,889,772 (90) | 214,150 (77)[a] | 54,835 (65)[abc] | 29,043 (72)[ab] | 130,272 (86)[a] | 1,675,622 (92) |
| **30-day outcomes (among survivors), n (%)** | | | | | | |
| Death in 30 days of discharge | 5,623 (0) | 1,542 (1)[a] | 580 (1)[abc] | 248 (1)[ab] | 714 (0)[a] | 4,081 (0) |
| Trajectory group for encounter with readmission within 30 days of discharge, n(%) | | | | | | |
| Persistent AKI without renal recovery | 23,347 (4) | 13,733 (17)[a] | 10,301 (49)[abc] | 895 (7)[a] | 2,537 (6)[a] | 9,614 (2) |
| Persistent AKI with renal recovery | 13,414 (2) | 7,988 (10)[a] | 965 (5)[abc] | 5,469 (40)[ab] | 1,554 (4)[a] | 5,426 (1) |
| Rapidly Reversed AKI | 45,584 (8) | 26,519 (34)[a] | 3,083 (15)[abc] | 1,680 (12)[ab] | 21,756 (50)[a] | 19,065 (4) |
| No AKI | 456,839 (85) | 30,505 (39)[a] | 6,795 (32)[abc] | 5,611 (41)[a] | 18,099 (41)[a] | 426,334 (93) |
| **Other complications during entire hospitalization** | | | | | | |
| Venous Thromboembolism, n (%) | 72,607 (3) | 15,706 (6)[a] | 4,802 (6)[abc] | 3,825 (9)[ab] | 7,079 (5)[a] | 56,901 (3) |
| Sepsis, n (%) | 129,683 (6) | 44,517 (16)[a] | 15,327 (18)[abc] | 10,055 (25)[ab] | 19,135 (13)[a] | 85,166 (5) |
| Cardiovascular complication, n (%) | 107,460 (5) | 36,092 (13)[a] | 11,773 (14)[abc] | 6,769 (17)[ab] | 17,550 (12)[a] | 71,368 (4) |
| Thirty-day mortality, n (%) | 15,294 (1) | 7,069 (3)[a] | 5,051 (6)[abc] | 660 (2)[ab] | 1,358 (1)[a] | 8,225 (0) |

| Variables | All Cohort (N=2,103,776) | AKI (N=276,987) | Persistent AKI without renal recovery (N=84,508) | Persistent AKI with renal recovery (N=40,527) | Rapidly Reversed AKI (N=151,952) | No AKI (N=1,826,789) |
|---|---|---|---|---|---|---|
| One-year mortality, n (%) | 49,094 (2) | 14,947 (5)[a] | 7,182 (8)[abc] | 2,295 (6)[ab] | 5,470 (4)[a] | 34,147 (2) |
| Three-year mortality, n (%) | 75,218 (4) | 20,183 (7)[a] | 8,359 (10)[abc] | 3,157 (8)[ab] | 8,667 (6)[a] | 55,035 (3) |

Abbreviations. SD, standard deviation; IQR, interquartile range; KRT, kidney replacement therapy; ICU, intensive care unit; LTAC, long-term acute care hospital; SNF, skilled nursing facility; NA, not applicable.

[a] $p<0.05$ compared to no AKI

[b] $p<0.05$ compared to Rapidly Reversed AKI

[c] $p<0.05$ compared to Persistent AKI with renal recovery

**Supplemental Table 12.** Association between AKI trajectory groups and hospital mortality in all cohort, ICU cohort, and non-ICU cohorts.

| | **Unadjusted Odds Ratio (95% Confidence Interval)** | **Model A Adjusted Odds Ratio (95% Confidence Interval)** | **Model B Adjusted Odds Ratio (95% Confidence Interval)** | **Model C Adjusted Odds Ratio (95% Confidence Interval)** |
|---|---|---|---|---|
| **All cohort** | | | | |
| No AKI | 1 | 1 | 1 | 1 |
| Rapidly Reversed AKI | 2.16 (0.04, 2.28) | 1.86 (1.76, 1.97) | 1.62 (1.53, 1.72) | 1.76 (1.67, 1.87) |
| Persistent AKI with Recovery | 5.30 (4.96, 5.66) | 4.39 (4.11, 4.70) | 3.11 (2.89, 3.34) | 3.57 (3.33, 3.82) |
| Persistent AKI without Recovery | 26.31 (25.51, 27.14) | 23.23 (22.51, 23.98) | 16.02 (15.35, 16.73) | 18.37 (17.72, 19.04) |
| **ICU cohort** | | | | |
| No AKI | 1 | 1 | 1 | 1 |
| Rapidly Reversed AKI | 1.15 (1.06, 1.24) | 1.08 (0.99, 1.17) | 1.04 (0.95, 1.13) | 1.05 (0.97, 1.14) |
| Persistent AKI with recovery | 1.62 (1.47, 1.78) | 1.50 (1.36, 1.66) | 1.34 (1.19, 1.49) | 1.33 (1.20, 1.47) |
| Persistent AKI without recovery | 19.39 (18.34, 20.49) | 18.39 (17.38, 19.45) | 16.12 (14.88, 17.46) | 15.57 (14.59, 16.62) |
| **Non-ICU cohort** | | | | |
| No AKI | 1 | 1 | 1 | 1 |
| Rapidly Reversed AKI | 1.93 (1.78, 2.09) | 1.66 (1.53, 1.80) | 1.46 (1.34, 1.59) | 1.59 (1.46, 1.72) |
| Persistent AKI with recovery | 1.66 (1.53, 1.80) | 4.36 (3.97, 4.79) | 3.16 (2.86, 3.50) | 3.65 (3.32, 4.03) |
| Persistent AKI without recovery | 25.14 (24.10, 26.22) | 21.86 (20.93, 22.82) | 15.44 (14.56, 16.37) | 17.93 (17.08, 18.83) |

Model A is adjusted for age, gender, ethnicity, and Charlson comorbidity score.
Model B is adjusted for age, gender, ethnicity, Charlson comorbidity score, and severe AKI (AKI stage≥2).
Model C is adjusted for age, gender, ethnicity, Charlson comorbidity score, and indicator of AKI stage 3.

**Supplemental Table 13.** Hazard ratios for all-cause mortality by AKI trajectories in all cohort, ICU cohort, and non-ICU cohorts.

| | **Unadjusted Hazard Ratio (95% Confidence Interval)** | **Model A Adjusted Hazard Ratio (95% Confidence Interval)[d]** | **Model B Adjusted Hazard Ratio (95% Confidence Interval)[d]** | **Model C Adjusted Hazard Ratio (95% Confidence Interval)[d]** |
|---|---|---|---|---|
| **All cohort** | | | | |
| No AKI | 1 | 1 | 1 | 1 |
| Rapidly Reversed AKI | 2.4 (2.3, 2.5) | 1.3 (1.2, 1.3) | 1.26 (1.22, 1.3) | 1.26 (1.22, 1.3) |
| Persistent AKI with Recovery | 4.2 (4.0, 4.4) | 1.3 (1.3, 1.4) | 1.35 (1.29, 1.4) | 1.34 (1.28, 1.4) |
| Persistent AKI without Recovery | 8.5 (8.3, 8.7) | 3.7 (3.6, 3.8) | 3.81 (3.68, 3.9) | 3.78 (3.68, 3.9) |
| C-index | 0.66 | 0.85 | 0.85 | 0.85 |
| **ICU cohort** | | | | |
| No AKI | 1 | 1 | 1 | 1 |
| Rapidly Reversed AKI | 1.5 (1.4, 1.6) | 1.15 (1.09, 1.22) | 1.29 (1.22, 1.37) | 1.30 (1.23, 1.37) |
| Persistent AKI with Recovery | 2.0 (1.9, 2.1) | 1.21 (1.13, 1.29) | 1.61 (1.50, 1.74) | 1.61 (1.51, 1.73) |
| Persistent AKI without Recovery | 7.5 (7.2, 7.8) | 4.62 (4.44, 4.81) | 5.94 (5.63, 6.27) | 5.89 (5.64, 6.16) |
| C-index | 0.67 | 0.78 | 0.75 | 0.75 |
| **Non-ICU cohort** | | | | |
| No AKI | 1 | 1 | 1 | 1 |
| Rapidly Reversed AKI | 2.2 (2.1, 2.3) | 1.3 (1.3, 1.4) | 1.5 (1.4, 1.6) | 1.5 (1.5, 1.6) |
| Persistent AKI with Recovery | 3.5 (3.3, 3.7) | 1.5 (1.4, 1.6) | 2.0 (1.9, 2.2) | 2.1 (2.0, 2.2) |
| Persistent AKI without Recovery | 6.5 (6.3, 6.7) | 3.1 (3.0, 3.2) | 4.0 (3.8, 4.1) | 4.1 (4.0, 4.2) |
| C-index | 0.63 | 0.81 | 0.78 | 0.78 |

Model A is adjusted for age, gender, ethnicity, and Charlson comorbidity score, need for mechanical ventilation and need for intensive care unit admission.
Model B is adjusted for variables as adjusted by Model A and severe AKI (AKI stage≥2).
Model C is adjusted for variables as adjusted by Model A and indicator of AKI stage 3.